\documentclass[twocolumn,superscriptaddress,pre]{revtex4-1}

\usepackage{graphicx}
\usepackage{amsmath,amssymb,amsfonts}

\newcommand{\Eq}[1]{Eq.~(\ref{eq:#1})}

\newcommand{\sol}{_\textrm{s}}  
\newcommand{\liq}{_\textrm{l}}  
\newcommand{\vap}{_\textrm{v}}  
\renewcommand{\sl}{_\textrm{sl}}  
\newcommand{\lv}{_\textrm{lv}}  
\newcommand{\sv}{_\textrm{sv}}  
\renewcommand{\sf}{\sl}  
\newcommand{\fv}{\lv}  
\newcommand{\hsf}{L\sf}  
\newcommand{\hfv}{L\fv}  
\newcommand{\hh}{h}  
\newcommand{\xx}{x}  

\newcommand{\mhsf}{\hat L\sf}  
\newcommand{\mhfv}{\hat L\fv}  
\newcommand{\mhh}{\hat h}  

\begin{document}

\title{How ice grows from premelting films and water droplets}

\author{David N. Sibley}
\affiliation{Department of Mathematical Sciences, Loughborough University,
Loughborough LE11 3TU, United Kingdom}

\author{Pablo Llombart}
\affiliation{Instituto de Qu\'{\i}mica F\'{\i}sica Rocasolano, CSIC, Calle
Serrano 119, 28006 Madrid, Spain}
\affiliation{Departamento de  Qu\'imica F\'isica (Unidad de I+D+i Asociada al
CSIC), Facultad de Ciencias Qu\'imicas, Universidad Complutense de Madrid,
28040, Spain}

\author{Eva G.  Noya}
\affiliation{Instituto de Qu\'{\i}mica F\'{\i}sica Rocasolano, CSIC, Calle
Serrano 119, 28006 Madrid, Spain}

\author{Andrew J. Archer}
\affiliation{Department of Mathematical Sciences, Loughborough University,
Loughborough LE11 3TU, United Kingdom}

\author{Luis G.~MacDowell}
\affiliation{Departamento de  Qu\'imica F\'isica (Unidad de I+D+i Asociada al
CSIC), Facultad de Ciencias Qu\'imicas, Universidad Complutense de Madrid,
28040, Spain}
\email{lgmac@quim.ucm.es}

\begin{abstract}
{
   Close to the triple point, the surface of ice is covered by a thin
   liquid layer {(so-called quasi-liquid layer)}
   which crucially impacts growth and melting rates.
   Experimental probes cannot observe the growth
   processes below this layer, and classical models of growth by vapor
   deposition do not account for the formation of {premelting} films.
   Here, we develop a mesoscopic model of liquid-film mediated ice growth,
   and identify the various resulting growth regimes.
   At low saturation, freezing proceeds 
   by terrace spreading, but the motion of the buried solid is conveyed through the liquid to
   the outer liquid-vapor interface. At higher saturations water droplets condense,
   a large crater  forms below, and freezing
   proceeds undetectably beneath the droplet.  
   Our approach is a general framework that naturally models freezing close
   to three phase coexistence and provides a first principle theory of ice
   growth and melting {which may prove useful} in the geosciences.
}
\end{abstract}

\maketitle

\section{Introduction}
\label{Intro}

The growth and melting of ice plays a crucial role
in numerous processes, from the precipitation of snowflakes
\cite{pruppacher10}, to glacier dynamics \cite{dash06}, scavenging of 
atmospheric gases \cite{abbat03} or climate change \cite{bartels-rausch13}. 
Yet, despite ice ubiquity both in large masses on the poles and as 
tiny crystals in the atmosphere, we still do not fully understand how ice 
actually grows (or melts) \cite{peter06,ball16,slater19,bonn20}.

Conflicting experimental measurements of ice growth
rates \cite{kobayashi67,lamb72,sei89,nelson98,libbrecht17} have 
been analyzed under a framework of classical crystal growth based 
on direct deposition from the vapor phase, followed by
the subsequent two dimensional migration of adatoms
onto surface kinks \cite{burton51}.
 However, the last two decades have
 witnessed great progress in the experimental characterization
 of the ice/vapor interface at equilibrium \cite{slater19}. Results
 from different experimental techniques \cite{elbaum93, wei01, bluhm02,
    sadtchenko02}, as well as computer
 simulations confirm that the surface disorder of ice grows steadily as the
 triple point is approached, and what is sometimes referred to as a
 `quasi-liquid layer' of  premelted
 ice is formed on its surface
 \cite{conde08,benet16,pickering18,qiu18,llombart20b}. 
Unfortunately, 
 classical growth models based on the terrace-ledge scenario do not account 
 for the impact of premelting films at all and attempts to
 incorporate this effect have met only limited success
 \cite{lamb74,kuroda82,neshyba16}.

Our current understanding of snow crystal growth illustrates this 
uncomfortable situation. The primary habit, 
or aspect ratio of
these familiar hexagonal crystallites can change dramatically with
small changes in temperature and 
saturation, from extremely elongated needle like crystals to almost
flat plate-like dendrites \cite{nakaya54}. But despite their variety and 
complexity, these shapes can be described
using phenomenological models with amazing accuracy, based on
just a number of parameters \cite{barret12,demange17}. Particularly, the
primary habit is dictated by a kinetic growth anisotropy
factor, describing the ratio of horizontal to vertical growth
rates \cite{demange17}. Unfortunately, the mapping  of this 
phenomenological  parameter
to the actual ambient conditions in the atmosphere, namely,
temperature and water saturation remains  a long standing
topic in crystal growth science \cite{kuroda82,nelson98,libbrecht17}. 
Accounting explicitly for the premelting layer appears an
essential requisite to unveil the dependence of growth
rates on ambient conditions.

 The difficulty to incorporate the role of premelting films on
 crystal growth theories is also encountered in many systems of
 interest in materials science \cite{pina98,deyoreo15,lutsko19},
 where the partially stable liquid phase can even 
 condense into  liquid droplets on the growing substrate \cite{elbaum93,
lazar05, murata16, jiang17} and change the mechanism of crystal
growth substantially.

The problem is akin to one encountered in the theory of wetting, where one studies
how a metastable liquid phase (say, water), adsorbs at the interface
between a solid substrate (ice) in contact with a vapor (water vapor)
as the liquid/vapor coexistence line  is traversed \cite{bonn01}.  
For an inert substrate, wetting 
is very well understood in terms of the underlying interface potential
$g( h )$ that measures the free energy of the adsorbed film as a
function of film thickness $ h $ \cite{schick90}. 
{Out of equilibrium, however, the substrate continually feeds from the adsorbed film at
   the expense of the mother phase, so it is debatable whether it is possible to
define meaningfully a film thickness and corresponding interface potential.}

Here, we combine state of the art computer simulations, equilibrium wetting
theory and thin-film modeling to describe the kinetics of the ice surface in the
vicinity of the triple point within a general framework for wetting on
reactive substrates.
Our results show that as the vapor saturation increases, 
ice first grows by terrace spreading below a premelting film 
with a well defined stationary thickness. At higher saturations, however,
the premelting layer thickness diverges, and growth actually proceeds
from below a bulk water phase. In between these two regimes,  at intermediate saturations, 
droplets condense on the ice surface and growth 
proceeds mainly under the droplets. The different regimes are separated by well defined
kinetic phase lines, whose location can be mapped to an underlying equilibrium
interface potential.

\section{Results}

\subsection{Interface potential for water on ice}

{

Most experiments in the literature report premelting layer thicknesses as a
function of temperature.  However, premelting can also be understood as
the condensation of water vapor onto the bulk ice surface. Viewed as an
adsorption problem, one sees that the layer thickness is
both a function of temperature and vapor pressure \cite{kuroda82}. 
Strictly, ice in contact with water vapor can 
only be in equilibrium along the sublimation line.
It follows that the
premelting thickness away from the sublimation line can
only be meaningfully characterized  for  small  deviations away from 
coexistence, where vapor condensation and freezing occur at exactly the same 
rate. Ice can then be out of equilibrium, while the premelting film remains 
in a stationary state of constant thickness \cite{kuroda90}. 
The failure to recognize
this important point is the source of much confusion in the literature 
and largely explains
why results for the premelting layer thickness differ by orders of magnitude
close to the triple point.

Here we show that an analysis of equilibrium surface fluctuations
of ice along the sublimation line can be exploited to calculate an
approximate interface potential for the premelting film. Input in
a suitable theory of crystal growth dynamics, this allows us to characterize 
the premelting layer thickness at arbitrary temperature and pressure.

To see this, we write the effective surface free energy per
unit surface area at solid/vapor coexistence as 
{$\omega( h ;T)=g( h ;T) - \Delta p_{\rm lv}(T) h $}, 
where {$\Delta p_{\rm lv}(T)$} is the pressure difference between the
liquid and vapor bulk phases at the solid/vapor coexistence chemical potential.
The free energy $\omega( h ;T)$ may be calculated over a limited
range of $ h $, by simulating  at solid/vapor equilibrium. During the course of the simulation, the film thickness 
fluctuates according
to a probability distribution $P( h ;T)$, which can be easily collected.
This can be used to obtain the free energy from the
standard fluctuation
formula {$\omega( h ;T)=-k_{\rm B}T\ln P( h ;T)$}, where {$k_{\rm B}$} is Boltzmann's 
constant \cite{macdowell06, hoyt09}. On the other hand,
{$\Delta p_{\rm lv}(T) $} is a purely bulk property and can be readily calculated
by thermodynamic integration from available data (see Methods and
Ref.\cite{llombart20}). With both $\omega( h ;T)$ and {$\Delta p_{\rm
lv}(T) $}
at hand, a batch of simulations along the sublimation line 
provide {$g( h ;T) = \omega( h ;T) + \Delta p_{\rm lv}(T) h $} for
a set of temperatures over a range of overlapping film thicknesses.
Since the interface potential is expected to exhibit only a small
temperature dependence, the set of piecewise functions $g( h ;T)$
at different temperatures may be combined to build a master
curve $g( h )$  over the whole range of
film thicknesses spanned in the temperature interval of the simulations
(see {Methods} and {Supplementary Note 1}).

{In principle, computer simulations of ice premelting are extremely challenging.
   The environment of a given molecule changes from solid to liquid and
   then to vapor over the scale of just one nanometer or less. The
   local polarization changes significantly across the interface, and therefore
   the average many body forces differ greatly depending on the local
   position. Such a complicated situation is best described with 
   electronic quantum-mechanical calculations, or explicit many body potentials
   \cite{gillian16,pham17}. Unfortunately, simulations with this level
   of detail for system sizes as large as required here appear unfeasible.
   Therefore we employ the TIP4P/Ice model \cite{abascal05}. Although this 
   is a point-charge non-polarizable potential, it
predicts} accurately both the solid/liquid and liquid/vapor surface tensions \cite{benet19}. 
Furthermore, in the range between 210~K to 271~K it produces film thicknesses that lie between
3 to 10~\AA, consistent with a growing body of evidence from
experimental probes \cite{bluhm02,sadtchenko02,mitsui19}. 

The results obtained with the TIP4P/Ice model for thicknesses up to one
nanometer are analysed as described above to produce the interface potential 
shown in Fig.~\ref{gh_sim}.

\begin{figure}[t]
\includegraphics[width=0.99\columnwidth]{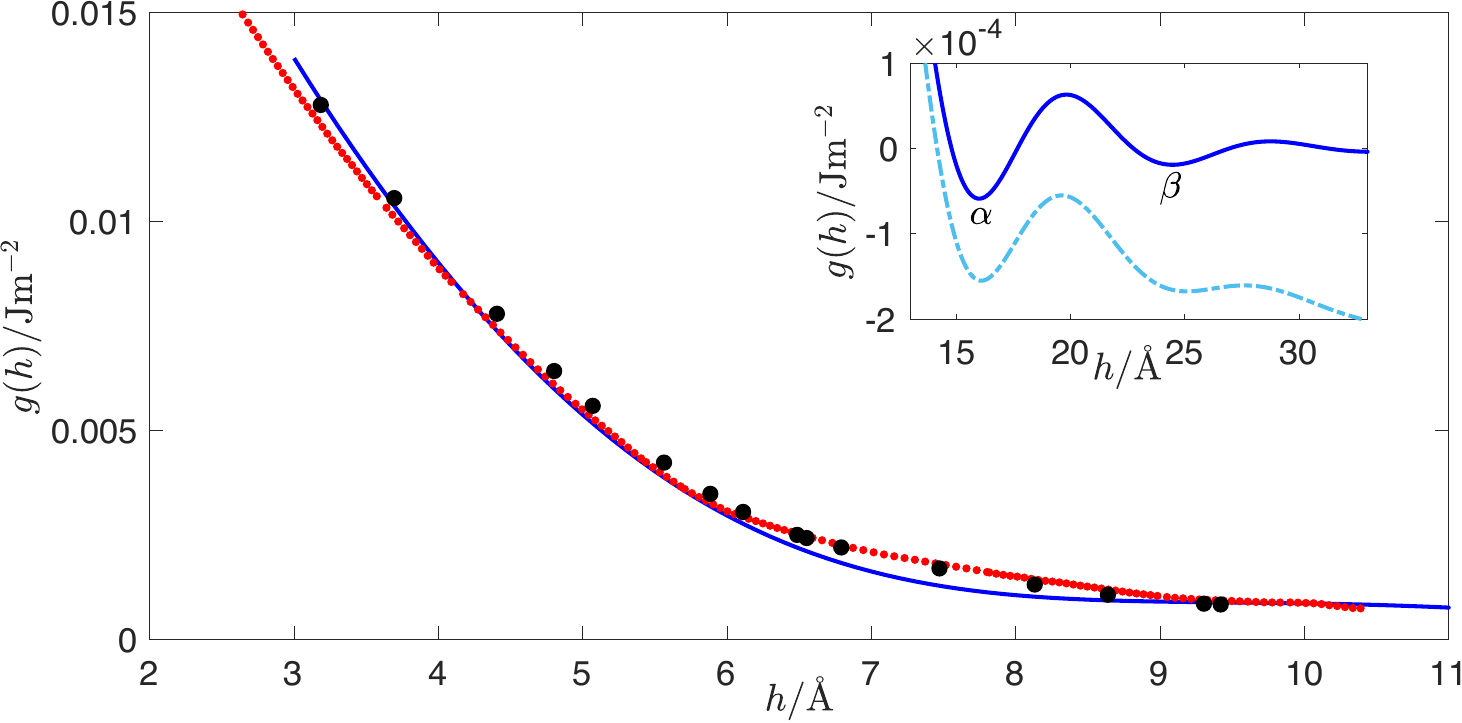}
\caption{
\label{gh_sim}
	   {\bf Interface potential for a water film adsorbed on ice as calculated from computer
	   simulations}. The small red circles are simulation results obtained from
{this work.} The larger black circles are results obtained
by integration of the related disjoining pressure as determined recently
\cite{llombart20}. The dark solid blue line is a fit to the simulation results, constrained to exhibit two minima. The inset shows details of the primary $\alpha$ and secondary $\beta$ minima, which are not visible on the scale of the main figure. {For an inert substrate, the $\beta$ state is stabilized at pressures
$\Delta p=46000$~Pa above liquid-vapor saturation
(dot-dashed light blue line)}.
}
\end{figure}

In practice, the equilibrium film thickness can grow well beyond
one nanometer as the triple point is approached, so that 
a complete model of the interface potential requires additional
input from theory and experiment.

Mean field liquid state theory shows that  a short range contribution of
the interface potential originating from molecular correlations in
the adsorbed film obeys the following equation \cite{chernov88, evans92,
henderson94, hughes17}:
{
\begin{equation}\label{eq:short-range}
   g_{\rm sr}( h ) = C_2 \exp(-\kappa_2  h ) - C_1 \exp(- \kappa_1  h )  \cos(q_0  h  +
\alpha),
\end{equation} 
}
where $C_i$ are positive constants, $\kappa_1$ and $\kappa_2$ are inverse decay
lengths (whichever is shorter is the inverse bulk correlation length), and
{$q_0 \approx 2\pi\,d_0^{-1}$}, where {$d_0$} is the molecular diameter. 

In practice, small amplitude capillary wave fluctuations  at
both the solid/liquid and liquid/vapor interfaces considerably
wash away the oscillatory behavior and renormalize the mean
field coefficients. 
Our computer simulations for the interface potential of the basal
face are consistent with this scenario:
fits  describe the simulations
accurately up to 10~\AA~, and then predict a fast decay with
very weak oscillations
of the sinusoidal term ({\emph{c.f.}}~\cite{llombart20}).


Additionally, there are algebraically decaying contributions to the interface potential which stem from the 
long range van der Waals interactions. These forces  originate from fluctuations of the electromagnetic field. 
Elbaum and Schick \cite{elbaum91b} parameterized the dielectric response of ice and water to numerically 
calculate these contributions with Dzyaloshinskii-Lifshitz-Pitaivesky theory. Following Ref.~\cite{macdowell19}, we show that the resulting crossover of retarded to non-retarded interactions is given accurately as
{
\begin{equation}\label{eq:crossover}
  g_{\rm vdw}( h )= -{B}{ h ^{-3}}\left [
		1 - f \exp(-a  h ) - (1-f) \exp(-b  h )
  \right ],
\end{equation} 
}
where $f$ is a parameter that accounts for the relative weight of infrared and ultraviolet contributions to the van der Waals forces, $a$ is a wavenumber in the ultraviolet region, while $b$ falls in the extreme-ultraviolet and accounts for the suppression of high frequency contributions (see 
{Supplementary Note 2} for further details).

The algebraic decay of the van der Waals forces
provides a negative contribution to the interface potential
and produces an absolute minimum at finite thickness 
\cite{elbaum91b,llombart20}.  This explains the observation
of water droplets formed on the ice surface just a few Kelvin away from
the triple point \cite{elbaum93,asakawa16,murata16}. The
droplets observed in experiment have a small contact angle of
$\theta=2^\circ$, which imply a shallow primary minimum  
with energy $\gamma_{\rm lv}(\cos\theta - 1) \sim {-}10^{-5}$~{ J m$^{-2}$ }.

Combining all this information, 
we obtain {$g( h )=g_{\rm sr}( h )+g_{\rm vdw}( h )$} and fit our
computer simulation results to this form, with $C_{\mathrm i}$, $\kappa_{\mathrm
i}$, $q_0$ and
$\alpha$ as fit parameters ({Supplementary Table~1} and {Supplementary
Note~3}). In fact, the simulation results can be fitted very
accurately to {$g_{\rm sr}( h )$} alone \cite{llombart20}, but extrapolation of the
simulation results to larger $ h $ is required to describe the behavior at
saturation. Therefore, in the parameter search we impose that $g( h )$ exhibits
minima at energies {$\sim-10^{-5}$~J m$^{-2}$}, as observed in experiment
\cite{murata16}. The constrained fit yields an interface potential in good
agreement with the available simulation data -- see Fig.~\ref{gh_sim}.
Consistent with expectations from renormalization theory, the shallow minima in
the interface potential are more widely spaced than one would expect from mean
field theory, located at $ h _{\alpha}=16.0$ \AA\ and $ h _{\beta}=24.5$ \AA. We
refer to these two as the $\alpha$- and $\beta$-minima, respectively, and this
interface potential provides a transition between a thin $\alpha$ and a thick
$\beta$ film at sufficiently large supersaturation as suggested in experiments
of ice premelting in the basal facet \cite{asakawa16,murata16}.

}

\subsection{Interface Hamiltonian}

\begin{figure}[t]
\includegraphics[width=0.99\columnwidth]{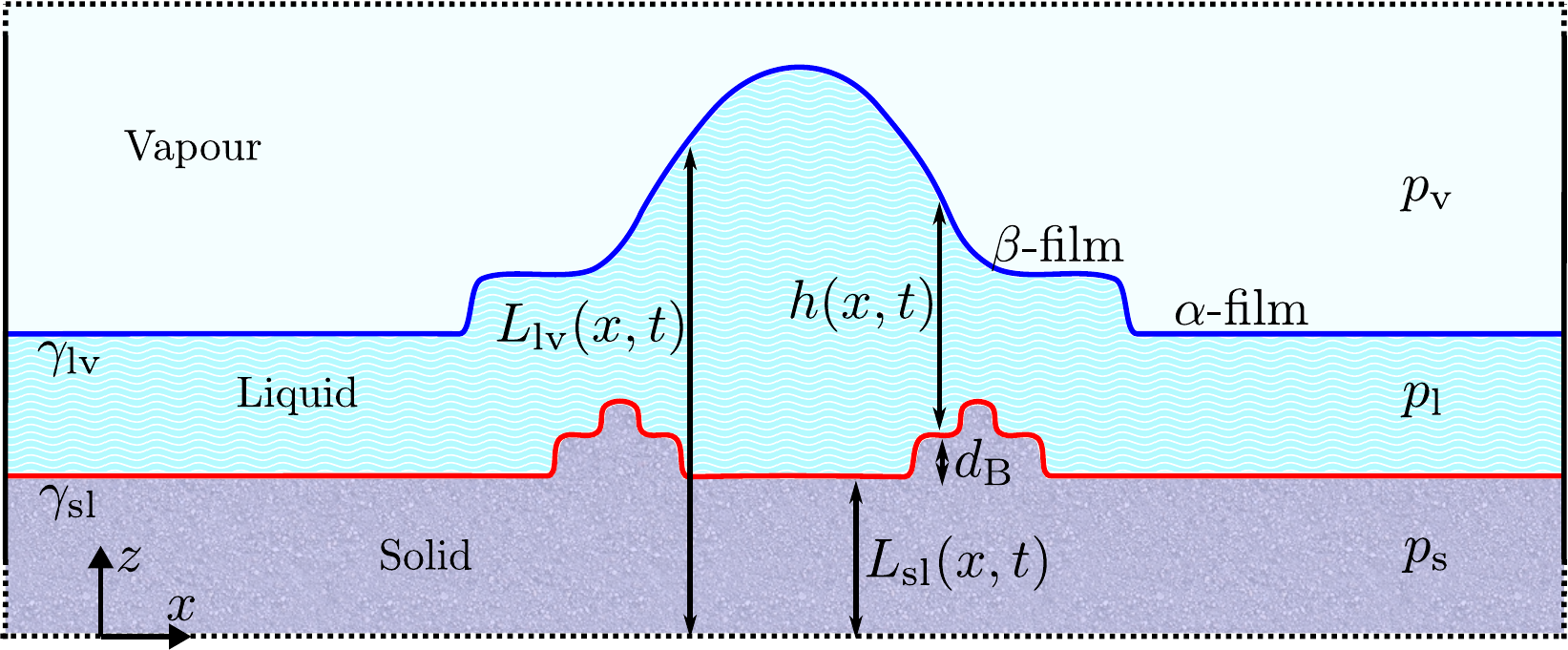}
\caption{
	   {\bf Illustration of a possible surface feature with annotations for
	   our two-dimensional gradient dynamics model setup}. Two evolving
	   interfaces are shown: the solid-liquid surface ({lower solid red line}) at reference
	   height {$z=L_{\rm sl}(x,t)$} and above the liquid-vapor interface
	   {(upper solid blue line)}
	   at reference height {$z=L_{\rm lv}(x,t)$}. The solid and vapour phases are modelled as extending infinitely below and above, respectively.
\label{IHscheme}
}
\end{figure}

The interface potential is adequate for describing equilibrium properties of homogeneous films. However, in order to account for droplets like that depicted in Fig.~\ref{IHscheme} and other such inhomogeneities, we must extend our description. Building on previous work \cite{benet16,benet19}, we begin by constructing a coarse-grained free energy (effective Hamiltonian) with all the required physics, consisting of a coupled sine-Gordon plus Capillary Wave (SG+CW) Hamiltonian with bulk fields,

{
\begin{align}\label{eq:sgcw}
\Omega  = & \int \left[\frac{\gamma_{\rm sl}}{2}(\nabla L_{\rm sl}
)^2+\frac{\gamma_{\rm lv}}{2}(\nabla L_{\rm lv})^2-u\cos(q_zL_{\rm sl}) \right. \nonumber\\
  &\qquad\qquad \left. + g(L_{\rm lv}-L_{\rm sl}) -\Delta p_{\rm sl}L_{\rm sl} -\Delta p_{\rm lv}L_{\rm lv} 
 \vphantom{\frac{\gamma_{\rm sl}}{2}} \right]{\rm d}\mathbf{ x }.
\end{align}
}

The first two terms account for the free energy cost to increase the surface
area of the solid/liquid and liquid/vapor surfaces in a long-wave approximation,
where 
{$L_{\rm sl}$} and {$L_{\rm lv}$} are {the height profiles of the two interfaces,} defined as the distances from the solid-liquid and liquid-vapor interfaces to an arbitrary reference  plane  that is parallel to the plane of the average ice surface
({\em c.f.} Fig.\ref{IHscheme}). Furthermore,
{$\gamma_{\rm sl}$} and {$\gamma_{\rm lv}$} are the solid/liquid interfacial
stiffness coefficient and the surface tension, respectively. The cosine term
accounts for the energy cost, $u$, 
to move the solid/liquid surface {$L_{\rm sl}$} away
from the equilibrium lattice spacing, as dictated by the wave-vector
{$q_z=2\pi\,d_{\rm B}^{-1}$}, where {$d_{\rm B}$} is the lattice spacing between ice bilayers at the
basal face. This simple model is known to describe adequately nucleated, spiral
and linear growth \cite{weeks79,bennett81,nozieres87,karma98}. The interface
potential coupling the two surfaces seeks to enforce the equilibrium thickness
of the premelting film {$ h=L_{\rm lv} -  L_{\rm sl} $}. The last two terms account
for the bulk energy of the system as measured relative to the (reservoir) vapor
phase with fixed chemical potential $\mu$, where {$\Delta p_{\rm sl}=
p_{\rm s}(\mu)-p_{\rm l}(\mu)$} is the pressure difference between the bulk solid and liquid
phases, while {$\Delta p_{\rm lv}=p_{\rm l}(\mu)-p_{\rm v}(\mu)$} is the pressure
difference between the bulk liquid and vapor phases. These two terms account for
the free energy change due to growth/melting of the solid phase at the expense
of the premelting film, and exchange of matter between the latter and the vapor
{\emph{via}} condensation/evaporation.

Note that the spectrum of equilibrium surface fluctuations of Eq.~(\ref{eq:sgcw}) can be
obtained exactly up to Gaussian renormalization \cite{benet16}. 
Accordingly, the parameters required in the theory can be obtained
in principle by requiring that the spectrum
of fluctuations from the theory match the results from experiments or simulations
\cite{benet19,llombart20b}. By virtue of this mapping, the input
of Eq.~(\ref{eq:sgcw}) is averaged over fluctuations, so that
$\Omega$ is to be interpreted as
a renormalized free energy, which incorporates consistently all
surface fluctuations in the scale of the parallel correlation length.

\subsection{Gradient driven dynamics}

The motion of the solid/vapor interface in the presence of a premelting film
necessitates us to account explicitly for the different dynamical processes
occurring at both the solid/liquid and liquid/vapor surfaces
\cite{lamb74,kuroda82,neshyba16}. On the one hand, {$L_{\rm sl}$} evolves
as a result of freezing/melting at the solid/liquid surface and on the other
hand, {$L_{\rm lv}$} evolves as a result of both the condensation/evaporation at the liquid/vapor surface and freezing/melting at the solid/liquid surface. Finally, we must also account for advective fluxes of the premelting film over the surface. In practice, since we are concerned only with small deviations away from equilibrium, we can assume the dynamics is mainly driven by free energy gradients with respect to the relevant order parameters \cite{Thiele_2010}. Accordingly, we treat the freezing/melting and condensation/evaporation in terms of non-conserved gradient dynamics, and the advective fluid dynamics of the premelting film using a thin-film (lubrication) approximation, whence

{
\begin{subequations}
\label{eq:dynsl}
\begin{equation}\label{eq:dynsl1}
\displaystyle   \frac{\partial L_{\rm sl}}{\partial t} = \displaystyle  
            -k_{\rm sl}\frac{\delta \Omega}{\delta L_{\rm sl}} \\
\end{equation}
\begin{equation}\label{eq:dynsl2}
\displaystyle  \frac{\partial L_{\rm lv}}{\partial t} =  \displaystyle  
\nabla\cdot
\left[\frac{ h ^3}{3\eta}\nabla\frac{\delta \Omega}{\delta L_{\rm lv}}\right] 
-k_{\rm lv}\frac{\delta \Omega}{\delta L_{\rm lv}}
+k_{\rm sl}
\frac{\Delta\rho}{\rho_{\rm l}}
\frac{\delta \Omega}{\delta L_{\rm sl}} 
\end{equation}
\end{subequations}
}

\noindent where {$k_{\rm sl}$} and {$k_{\rm lv}$} are kinetic growth coefficients that determine the rate
of crystallization and condensation at the solid/liquid and liquid/vapor
surfaces, respectively, $\eta$ is the viscosity in the liquid film and
{$\Delta\rho=\rho_{\rm s}-\rho_{\rm l}$}, where {$\rho_{\rm s}$} and
{$\rho_{\rm l}$} are the
densities of the solid and liquid, respectively. 
Models with some similar
features were developed in Refs.~\cite{pototsky05,yochelis07,Thiele_2010}.

Notice that the deterministic dynamics given by Eq.~(\ref{eq:dynsl}) is driven
by the renormalized free energy, Eq.~(\ref{eq:sgcw}). Accordingly,
the equation accounts for stochastic fluctuations implicitly, and it may 
be interpreted as dictating the evolution
of the film profiles averaged over all possible random trajectories
\cite{archer04}.
Alternatively, replacing
the renormalized free energy by a mean field Hamiltonian 
one can assume the
above result describes the most likely path of the system \cite{lutsko19}. 
When the fluctuations are small, the coarse grained Hamiltonian
and the renormalized free energy do not differ significantly, and
the evolution of the average trajectory becomes the same as
the most likely path, as expected in mean field theory. 
In the {Supplementary Note 4} we provide an extended discussion on this
issue and  show that Eq.~(\ref{eq:dynsl}) may be derived from
a fully stochastic driven dynamics of the mean field Hamiltonian.


\subsection{Kinetic phase diagram}

The time evolution predicted by
Eq.~(\ref{eq:sgcw}-\ref{eq:dynsl}) is extremely rich and varied
and the full range can only be obtained numerically. However, if we assume that
the surface is on average flat, then we obtain equations that enable us to
predict the outcome of the numerical simulations and determine an accurate 
kinetic phase diagram. Coarse-graining the evolution over the time period
required to form a single new plane of the crystal, we replace the time
derivatives of {$L_{\rm sl}$} and {$L_{\rm lv}$} by their average values (denoted as $\langle
\cdot \rangle$), yielding a rate law for continuous growth ({Supplementary
Note 5}):
{
\begin{subequations}
\label{eq:av_growth}
\begin{equation}
\label{eq:av_growth1}
   \displaystyle   
\langle{\partial_t L_{\rm sl}}\rangle = \pm k_{\rm sl} \sqrt{\phi_{\rm sl}^2-w^2}  
\end{equation}
\begin{equation}
\label{eq:av_growth2}
   \displaystyle   
\langle{\partial_t L_{\rm lv}}\rangle = \displaystyle  
 k_{\rm lv} \phi_{\rm lv} -({\Delta\rho}/{\rho_{\rm l}})\langle{\partial_t L_{\rm sl}}\rangle
\end{equation}
\end{subequations}
}
where $w=q_z u$, {$\phi_{\rm sl}=\Delta p_{\rm sl} -\Pi$},
{$\phi_{\rm lv}=\Delta p_{\rm lv} +\Pi$} and $\Pi(h)=-\partial_h g(h)$ is the disjoining pressure. 
In Eq.~\ref{eq:av_growth1}, the plus sign corresponds to freezing (
{$\phi_{\rm sl}>0$}), 
while the minus sign corresponds to sublimation ( {$\phi_{\rm sl}<0$}).

Despite the coarse graining, Eq.~(\ref{eq:av_growth}) predicts a complex dynamics
in very good agreement with the numerical solutions of Eq.~(\ref{eq:dynsl}) (see below).

Firstly, for points in the temperature-pressure plane where {$\phi_{\rm sl}^2<w^2$},
the crystal surface is pinned by the bulk crystal field and remains
smooth. Within this region, continuous growth cannot occur. Instead,
the loci of points obeying {$\phi_{\rm sl}^2=w^2$}
encloses a region of activated growth, where the crystal front advances 
{\emph{via}} nucleation and spread of new terraces \cite{bennett81,nozieres87}. 

For state points where {$\phi_{\rm sl}^2>w^2$}, the thermodynamic driving force
becomes larger than the pining field. The surface then undergoes kinetic
roughening, and growth can proceed continuously. The growth of the premelting
film thickness may be found by subtracting the growth rate of
{$\langle\partial_tL_{\rm sl}\rangle$} from that of {$\langle\partial_t L_{\rm lv}\rangle$}, 
yielding:
{
\begin{equation}\label{eq:sfl}
   \displaystyle \left \langle\frac{\partial  h }{\partial t} \right \rangle  =
  \displaystyle  k_{\rm lv} \phi_{\rm lv} \mp \frac{\rho_{\rm s}}{\rho_{\rm l}} k_{\rm sl}
\sqrt{\phi_{\rm sl}^2-w^2}.
\end{equation}
}
In practice, we are interested in mapping the phase diagram for
quasi-stationary states, where the solid and liquid phases grow at the
same rate, so that the premelting film thickness
remains constant, i.e. such that $ \langle\frac{\partial  h }{\partial t}\rangle = 0$
\cite{kuroda82,neshyba16}.
Solving for this equality provides a condition for the film thickness
as a function of pressure and temperature, which is conveniently written as:
{
\begin{equation}\label{eq:keq}
   \Pi( h ) = -\Delta p_{\rm k}(p_v,T),
\end{equation} 
}
where $\Pi( h )$ is the disjoining pressure, while {$\Delta p_{\rm k}(p_v,T)$}
is a function of the ambient conditions but 
depends parametrically also on the growth mechanism and rate constants (See
{Supplementary Note 5}).

To illustrate the significance of this equation, consider the simple case of a rough surface, i.e. such that 
 $w=0$. Then, solving Eq.~(\ref{eq:sfl}) for stationarity,
readily yields Eq.~(\ref{eq:keq}), with the kinetic overpressure given in the simple
form:
{
\begin{equation}\label{eq:stationarity1}
   \Delta p_{\rm k}(p_v,T) = \frac{\rho_{\rm s} k_{\rm sl}}{\rho_{\rm s} k_{\rm sl}+\rho_{\rm l} k_{\rm lv}}
                     \Delta p_{\rm sl}  -
          \frac{\rho_{\rm l} k_{\rm lv}}{\rho_{\rm s} k_{\rm sl}+\rho_{\rm l} k_{\rm lv}}
                     \Delta p_{\rm lv}
\end{equation}
}
Notice that {$\Delta p_{\rm sl}$} and {$\Delta p_{\rm lv}$} are purely bulk quantities that
only depend on the imposed thermodynamic conditions of the system, and
convey the state dependent information to the kinetic overpressure 
({Supplementary Table~2} and {Supplementary Note 6}). 
 In the limiting case where the
 substrate is strictly inert, {$k_{\rm sl}=0$}, then Eq.~(\ref{eq:stationarity1}) 
 becomes {$\Pi( h ) = - \Delta p_{\rm lv}$} exactly, which is the Derjaguin 
condition for the equilibrium
film thickness on inert substrates. This is very convenient, because we can then
predict the outcome of the non-equilibrium dynamics by analogy with the 
behavior of  equilibrium films on inert substrates, albeit with
the effective overpressure {$\Delta p_{\rm k}$} replacing {$\Delta
p_{\rm lv}$}. 
Likewise, one sees that an effective  interface potential {$\omega_{\rm k}( h ) =
g( h ) - \Delta p_{\rm k}  h $} determines the dynamics of the system in the
quasi-stationary regime.

\begin{figure*}[t]
\includegraphics[width=0.99\textwidth]{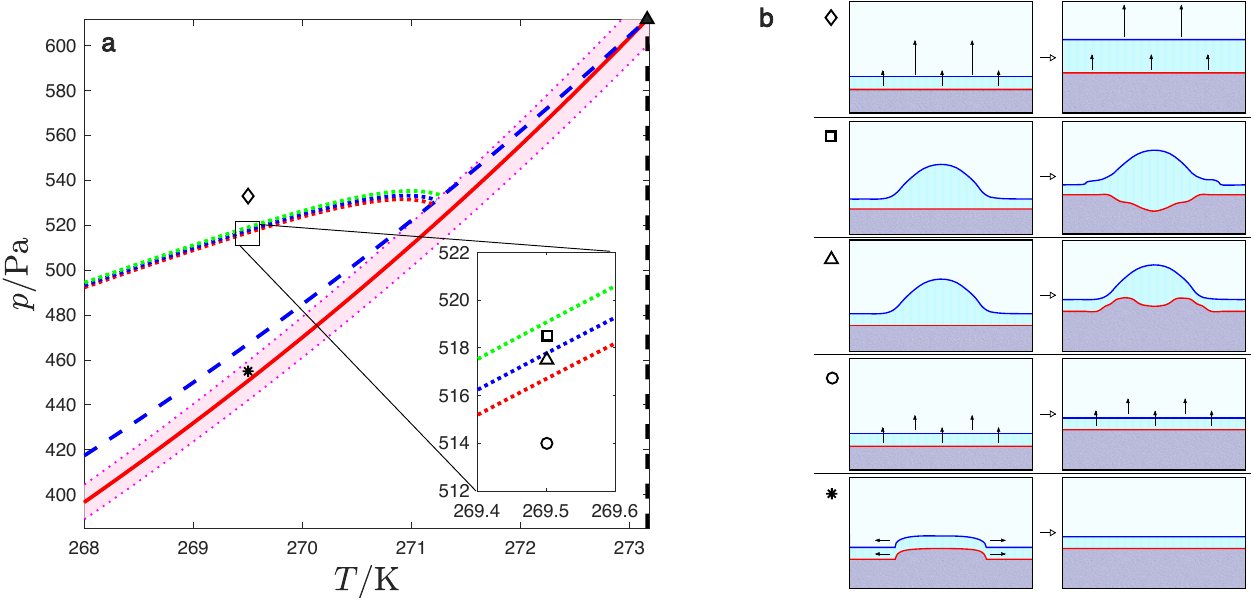}
\caption{
\label{fig:phaseD}
     {\bf Kinetic phase diagram for ice crystal growth.}
  {Panel \textbf{a} shows the equilibrium phase diagram and kinetic phase lines.}  The red solid line  is the sublimation line, whereas the dashed lines are metastable prolongations of the 
vaporization (blue) and melting (black) lines. The {filled triangle ($\blacktriangle$)} indicates
the triple point where these lines meet. The remaining features describe the outcome of the dynamics.  The shaded area designates the region of activated growth.  The dotted lines are kinetic phase lines corresponding 
to kinetic coexistence (red dotted), kinetic $\alpha\to\beta$ transition (blue dotted) and kinetic spinodal (green dotted) lines as explained in the text.
{Panel \textbf{b} shows sketches with the dynamics observed in different points of the phase diagram as indicated with the corresponding symbols.} 
The colored lines describe the ice/liquid (red) and the
liquid/vapor (blue) surfaces enclosing the premelting film.  The black arrows show the direction of preferential growth.
 {At the point marked by an asterisk ($*$),}
in the region of activated {dynamics,
growth proceeds} by horizontal translation of nucleated terraces.  
At points such as {that marked by a circle ($\circ$),} above the region of activated {dynamics,
growth can occur} continuously without activation in a steady state of constant film
thickness. {At points such as that marked by the open triangle ($\triangle$)},
above the kinetic coexistence line, 
droplets can condense and are stabilized transiently with a crater
growing inside. {At points such as that marked by a square ($\square$),} beyond the $\alpha\to\beta$ line,
films in the $\beta$-thick state can be stabilized
transiently and form at the rim of the droplet. At higher pressures,
past the kinetic spinodal line (green dotted), {such as the point marked with a lozenge ($\lozenge$)}, 
the crystal growth rate can no longer match the condensation rate,
and the film thickness diverges. The detailed dynamics corresponding
to symbols in the phase diagram is illustrated in Figures 4 {and 5}.
}
\end{figure*}

This allows us to determine the kinetic phase diagram, 
identifying the regions in 
$(p,T)$ space where the different outcomes of the interfacial wetting 
dynamics is to be expected (Fig.~3). 
In particular, we identify three significant kinetic phase lines:
\begin{itemize}
\item
The line of kinetic coexistence (dotted-red line in Fig.~\ref{fig:phaseD})
occurs when {$\Delta p_{\rm k}=0$}. The location of this line 
can be obtained from Eq.~(\ref{eq:keq}), for
the choice $\Pi( h )=0$. States above this line have stationary
film thickness consistent with $\Pi( h )<0$ and are effectively 
oversaturated. Accordingly, the Laplace condition for
droplet formation is met for the first time, and droplets can
be stabilized transiently.  However, this occurs
well above the liquid-vapor coexistence line, and explains why
droplets reported in experiment 
are formed only above the condensation line
\cite{asakawa16,murata16}. 
\item The line of $\alpha\to\beta$ kinetic transition (dotted-blue line in Fig.~\ref{fig:phaseD}).
   At sufficiently high saturation, the linear term in {$\omega_{\rm k}(h)$}
   stabilizes the $\beta$ state transiently, and it is possible to
   observe the coexistence between $\alpha$ and $\beta$ states that
   has been reported in experiment \cite{asakawa16,murata16}.
   The line where the condition is met is obtained by solving  a double
   tangent construct as in usual wetting phase diagrams ({Supplementary Note 5}). 
\item The kinetic spinodal line (dotted-green line in Fig.~\ref{fig:phaseD}),
   which occurs when {$\Delta p_{\rm k}=-\Pi_{\rm spin}$}, with
   {$\Pi_{\rm spin}$} the value at which the interface potential $g( h )$ predicts
   that the liquid/vapor interface {$L_{\rm lv}$} becomes linearly unstable, i.e.\ has a
   spinodal. This condition leads to a line {$p_{\rm spin}(T)$} that can be obtained
   from Eq.~(\ref{eq:keq}), for the choice {$\Pi=\Pi_{\rm spin}$}.  Crossing this line signals
  the region of the temperature plane where ice crystal growth cannot match
  the rate of vapor condensation, and the premelting film thickness
diverges.
\end{itemize}
The slope of the kinetic coexistence lines is dictated by the ratio of
{$k_{\rm sl}$} to {$k_{\rm lv}$}, while the separation between kinetic phase lines is dictated by the
depth and free energy separation between the minima.

\begin{figure*}[t]
	\includegraphics[width=0.99\textwidth]{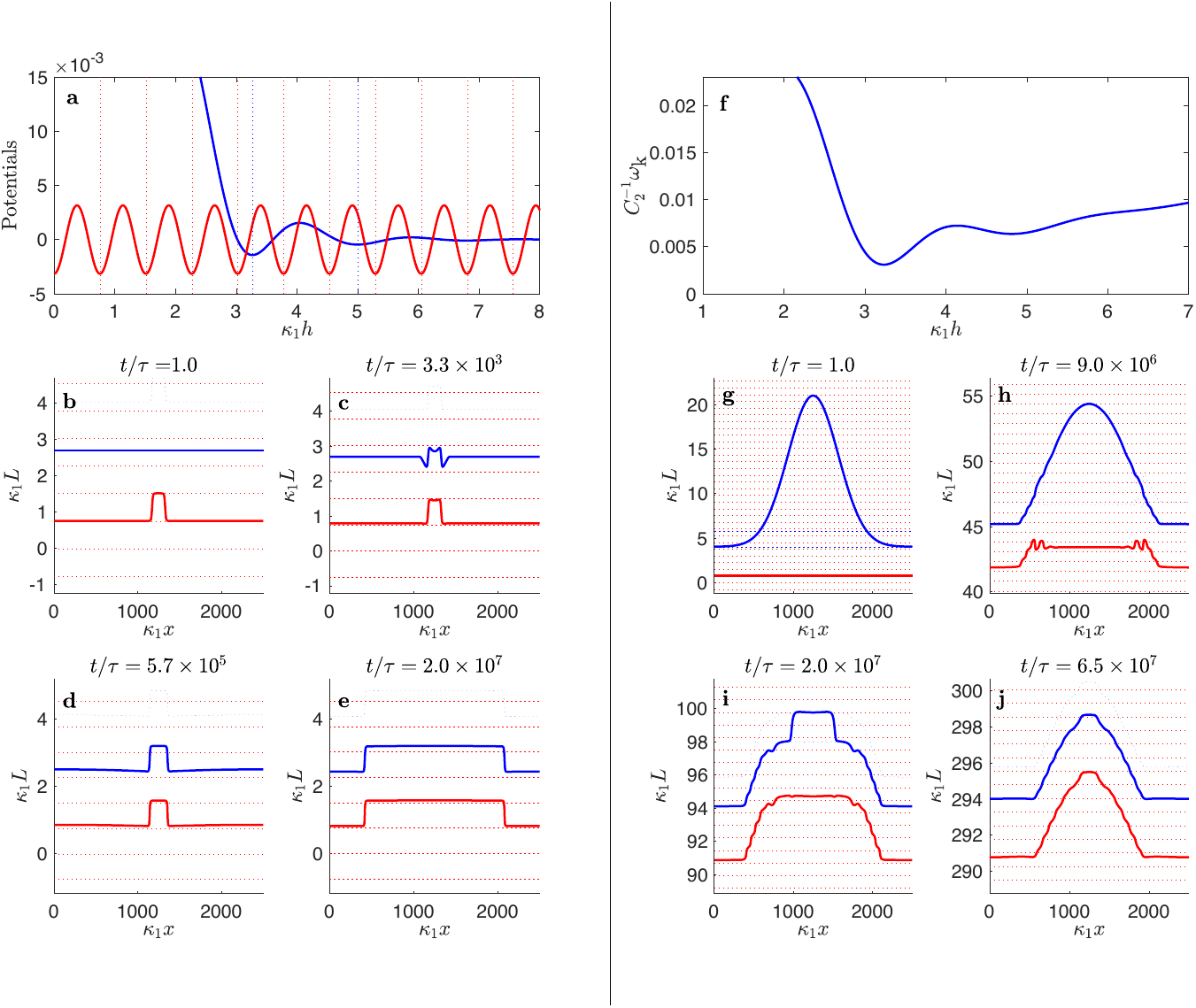}
\caption{
\label{snapshots1}
	{\textbf{Surface dynamics  below the kinetic coexistence line.} 
	Panels \textbf{a} and \textbf{f} show the effective potentials for state points
   depicted as an asterisk (*) and a circle ($\circ$) in Fig.~\ref{fig:phaseD}, respectively.} 
   {
   	Panels \textbf{b}-\textbf{e} and \textbf{g}-\textbf{j} } show
the corresponding solid/liquid and liquid/vapor surfaces at significant
milestones in their time evolution (solid red and blue lines, respectively). 
The dashed red lines indicate the surface location for fully formed ice bilayers, and dashed blue
lines show the  heights of a premelting film at the $\alpha$ or
$\beta$ minima, as a guide to the eye. 
{Panels \textbf{a}-\textbf{e} illustrate the} evolution in the
nucleated regime at $(p,T)=(455~Pa,269.5~K)$ ({marked as an
asterisk}  ($*$) in Fig.~\ref{fig:phaseD}). {Panel {\bf a} shows the sine-Gordon and interface potential which dictate the surface dynamics.}  
A small terrace nucleated on the solid/liquid surface ({panel {\bf b}}) triggers the formation of a similar terrace on the liquid/vapor surface
   {(panel {\bf c})} and then spreads horizontally ({panels \textbf{d}-\textbf{e}}). 
	{Once the surface has flattened, further growth
	is not possible until a new terrace is nucleated (Supplementary Movie 1)}. 
   {Panels \textbf{f}-\textbf{j} illustrate the evolution of}
a droplet quenched to a pressure just below the kinetic liquid-vapor
coexistence line at $(p,T)=(514~Pa,269.5~K)$ ({shown as a circle} ($\circ$) 
in Fig.~\ref{fig:phaseD}). {The effective free energy,  $\omega_{\rm k}( h)$, 
   (panel {\bf f}) inhibits the growth of liquid wetting films.  A
   droplet  (panel {\bf g}) triggers the formation of a terrace 
   at the rim, which then spreads inside (panel {\bf h})  and grows to fill the 
droplet completely (panels \textbf{i}-\textbf{j}). Subsequent growth occurs in a
quasi-stationary state of constant film thickness (Supplementary Movie 2) }.
}
\end{figure*}

Using gas kinetic theory, crystal growth theory, and literature data for water
and ice 
we estimate the model parameters
{$k_{\rm sl}$}, {$k_{\rm lv}$}, $w$, $\eta$, {$\gamma_{\rm sl}$},
{$\gamma_{\rm lv}$}, {$\Delta p_{\rm sl}$} and 
{$\Delta p_{\rm lv}$}
for the basal surface of ice ({Supplementary Table~3} and
{Supplementary Note~7}).  
These data, combined with the interface potential $g( h )$ from computer
simulations, allows us to draw the kinetic phase diagram depicted in
Fig.~\ref{fig:phaseD}. The shaded area surrounding the sublimation line is the
region where crystal growth is a slow activated process, only proceeding
{\emph{via}}
step nucleation and growth.
In the absence of any impurities to speed up the nucleation, in
this regime the substrate is effectively unreactive for time scales smaller than
the inverse nucleation rate, and behaves {as dictated by} the equilibrium interface
potential, Fig.~\ref{snapshots1}-a. 
In practice, the experimental systems reported in Ref.\cite{murata16} contain dislocations, 
so the crystal freezes by spiral growth and the region of unreactive wetting
shown in Fig.~\ref{fig:phaseD} for the SG+CW model is not observed.
The significance of this change in the growth mechanism can be illustrated
by setting $w=0$. In this case, the region of activated growth is removed
altogether, growth proceeds continuously and the kinetic phase lines all meet the solid/liquid coexistence
line as they approach the triple point ({Supplementary Figure 1}).
This regime is also relevant for the prism plane above its roughening transition
at about 269~K.

\subsection{Interface dynamics}

\begin{figure*}[t]	
	\includegraphics[width=0.99\textwidth]{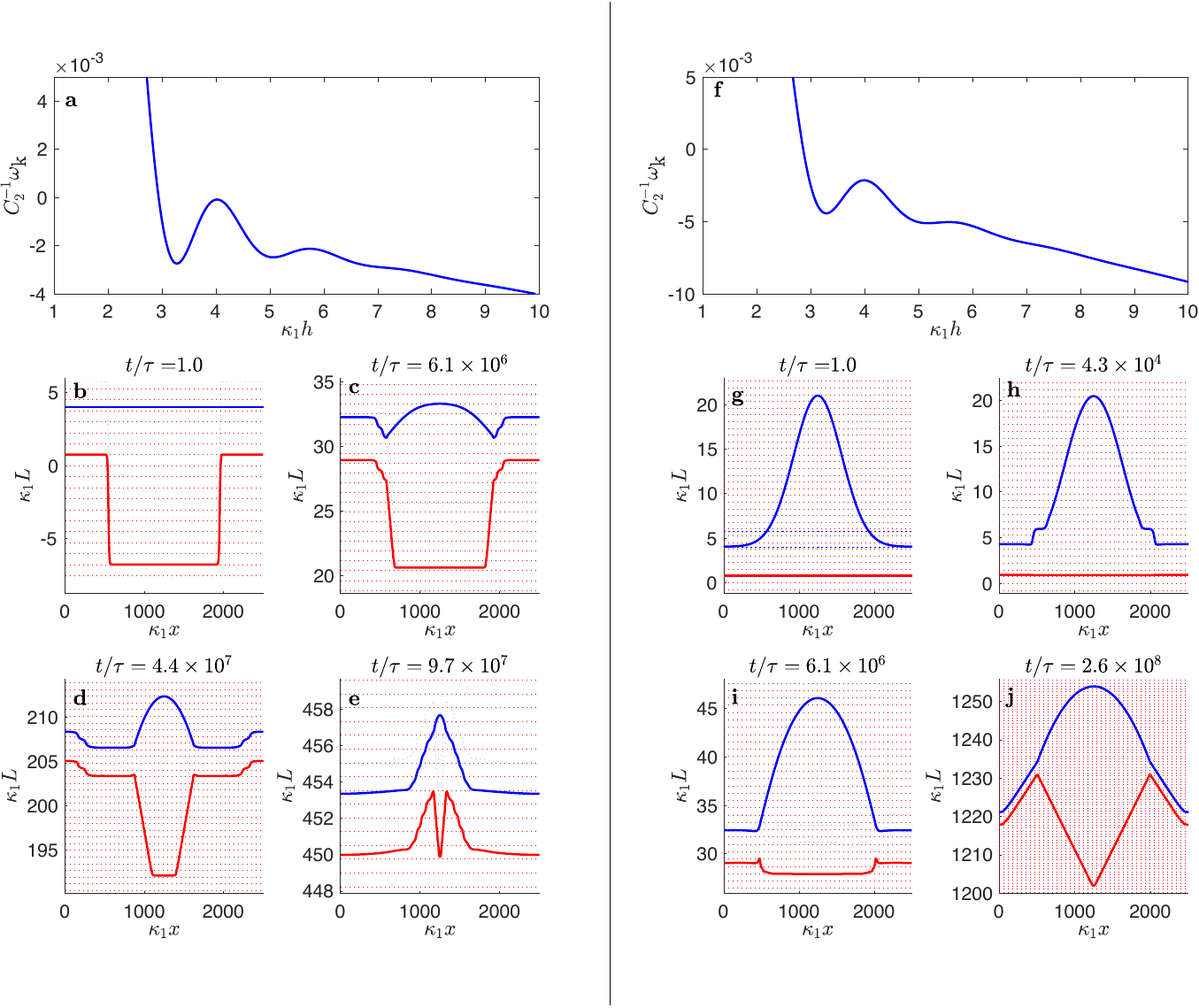}
\caption{
\label{snapshots2}
	 {\bf Surface dynamics  above the kinetic coexistence line.} 
	{   Panels {\bf a} and {\bf f} show the effective free energies, $\omega_{\rm k}( h)$ 
	   that drive the time evolution of the ice surface at state points
	   depicted as a triangle ($\triangle$) and a square ($\square$) in Fig.~\ref{fig:phaseD}, respectively.
Panels \textbf{b}-\textbf{e} and \textbf{g}-\textbf{j} show solid/liquid and liquid/vapor surfaces at
significant milestones of the dynamics as described in Fig.~4.
Panels \textbf{a}-\textbf{e} display the evolution of a surface defect at state point
$(p,T)=(517.5~Pa,269.5~K)$ (shown as a triangle ($\triangle$) in Fig.~\ref{fig:phaseD}).
The growth of a thick wetting film is now favorable, as illustrated
by the negative slope of the effective free energy in panel {\bf a}.
A defect on the solid/liquid surface (panel {\bf b}) triggers the formation
of a liquid droplet (panel {\bf c}). Ice then grows inside the droplet,
forming a large crater (panels \textbf{d}-\textbf{e}) which vanishes eventually when the
ice surface catches up with the liquid droplet and attains a stationary
premelting layer thickness (Supplementary Movie 3). 
Panels \textbf{f}-\textbf{j} display the evolution of a droplet at 
$(p,T) = (518.5~Pa,269.5~K)$ above the kinetic $\alpha\to\beta$ transition
line (shown as a square ($\square$ ) in  Fig.~\ref{fig:phaseD}).
Here, the $\beta$ state has lower free energy than the $\alpha$ state,
as illustrated in panel {\bf f}.  During the time evolution of
a droplet (panel {\bf g}), a thick film of $\beta$ thickness
forms at the rim transiently (panel {\bf h}), then the droplet evolves
a crater  (panels \textbf{i}-\textbf{j}) as the ice surface catches up with
the droplet (Supplementary Movie 4). 
}
}	
\end{figure*}

An extensive set of numerical simulations performed over a wide range
   of the $p-T$ plane and initial conditions confirms that the outcome of
   the dynamics is in excellent agreement with expectations from the kinetic phase diagram
   of Fig.~\ref{fig:phaseD}.
   Here we report results performed for the basal surface 
   at $T=269.5$~K and varying vapor pressure. Results are reported
   in reduced units of the model parameters, with  $\kappa_1^{-1}\approx
0.49$~nm for the length scale and 
{ $\tau = 3\eta\,(\kappa_1 \gamma_{\rm lv})^{-1}\approx 0.11$ }~ns 
for the time scale.

 First we consider a state very near
   the sublimation line, where the system is found in the region of
   activated growth, and water vapor freezes by a growth and spread mechanism. 
   The premelting film here is virtually in equilibrium for the time scale
   of the simulation, and adopts a thickness very much equal to the value
at the absolute
minimum of $g( h )$, or $\alpha$ state.
In our simulations (Fig.~\ref{snapshots1}(\textbf{b}-\textbf{e}) and {Supplementary Movie 1}), an initial terrace
mimicking a local defect on the solid/liquid surface {$L_{\rm sl}$}, not observable by
optical means, triggers the formation of a corresponding terrace on the
liquid/vapor surface {$L_{\rm lv}$}, with a step height equal to the solid lattice
spacing. Crystal growth then proceeds by the spreading of the terrace, and the
horizontal motion of  the solid phase is conveyed to the external
liquid/vapor surface. {This motion can  be observed directly by 
   confocal microscopy, but of course, does not imply the absence
of a disordered premelting film 
({ \emph{c.f.}}~Fig.~5 in Ref.~\cite{asakawa16} or Ref.~\cite{murata16})}. 
Once the new full crystal lattice plane is formed, growth becomes stuck
again until a new critical nucleus is formed stochastically.

Crossing the line of nucleated growth towards higher saturation, such that
{$\phi_{\rm sl}>w$}, the thermodynamic driving force  is large enough
to beat the bulk crystal field, and growth then occurs without activation, as in
a kinetically rough surface  \cite{weeks79,nozieres87}. However, if
{$\phi_{\rm sl}$}
is only marginally larger than $w$, the process occurs in a stepwise fashion,
occurring with large time intervals of no growth, followed by height increments
equal to the lattice spacing $d_B$ in a short time \cite{neshyba16}. On further increasing
{$\phi_{\rm sl}$}, crystal growth then proceeds in a truly quasi-stationary manner
while the premelting film thickness remains constant, consistent with
Eq.~(\ref{eq:keq}).

Interestingly, traversing the metastable prolongation of the liquid-vapor
coexistence line does not change the growth behavior in any significant
way. Although
{$\Delta p_{\rm lv}$} is now positive, {$\Delta p_{\rm k}$} is still negative, so the
thickening of $ h $ is still uphill in the effective free energy
{ $\omega_{\rm k}( h )$}: i.e. the system behaves as if it is effectively undersaturated
with respect to liquid-vapor coexistence and the vapor/liquid interface cannot
advance faster than the crystal/liquid interface (\emph{c.f.}~Fig.4-\textbf{f}). 
For a purely flat interface, the stationary film thickness here is
   therefore somewhat smaller than that found at the sublimation line, but
still remains confined within the $\alpha$ state of the interface potential
(See Fig.~\ref{snapshots1}-{\textbf{f}}).
A liquid droplet quenched to this region of the kinetic phase diagram is never
stable -- see Fig.~\ref{snapshots1}({\textbf{g}}-{\textbf{j}}) and {Supplementary Movie 2}. Instead, at the contact
line of the droplet, terrace formation on the ice is triggered by the action of
the disjoining pressure. The crystal then grows and the droplet flattens out, in
order to reach a quasi-equilibrium film thickness consistent with
Eq.~(\ref{eq:keq}). As a
transient during the process, the premelting film thickness $ h $ can be stable
in the $\beta$ film state, reminiscent of the `sunny side up' states observed in
experiment \cite{murata16}. Subsequently, the droplet disappears, leaving an
Aztec pyramid shaped solid surface that is covered by an $\alpha$-thick film.
Finally, the inhomogeneity completely
disappears, and growth proceeds in a strictly quasi-stationary manner with a
flat surface. {Notice that during the relaxation process, the droplet
   is lifted upwards, as a result of the continuous ice growth occurring below.
   Indeed, comparing Fig.~\ref{snapshots1}-{\textbf{g}} with
Fig.~\ref{snapshots1}-{\textbf{j}},
   we find that well before the inhomogeneity is washed out, the
   ice surface grows by about 290 full lattice spacings, at a rate
   consistent with Eq.~(\ref{eq:av_growth}). This shows
   that the relevant relaxation time for large inhomogeneities is far
larger than the coarse graining time scale used to obtain the 
average growth rate law.}

The situation changes significantly when saturation is raised above the kinetic liquid-vapor coexistence line, where 
{$\Delta p_{\rm k}>0$}. 
For thick enough films, $ h $ can now move downhill in the effective surface
free energy ({Fig.~5-\textbf{a}}).
In this regime, small fluctuations or crystal defects that locally increase the
film thickness beyond the spinodal thickness of $g( h )$ trigger the formation
of large liquid droplets on top of the premelting film, as observed in
experiments---see {Fig.~\ref{snapshots2}(\textbf{b-e})} and {Supplementary Movie 3};
{\emph{c.f.}}~Fig.~1-D from \cite{murata16}. Essentially, when
{$\Delta p_{\rm k}>0$ }
the liquid pressure is large enough to sustain the tension of the droplet
surface. However, the droplet cannot be fully stable here, since the system is
open. The fastest way to decrease the overall free energy while the solid phase
grows is to form a large crater below the droplet and then for the two
interfaces to separate. Likewise, a droplet quenched to this region behaves
initially as described above for droplets below the kinetic liquid-vapor
coexistence. The difference is that once a few terraces have been formed at the
rim, the crystal grows thereon inside the droplet towards its center by creating
a premelting film of $\alpha$ thickness, without the droplet curvature
flattening out ({Supplementary Figure.~2} and {Supplementary Movie~5}). As growth proceeds, the interface profiles take a transient shape like that of droplets on soft substrates \cite{style13, andreotti20}, with the solid surface growing higher in the contact line region. A crater develops, but is then filled by the growing solid, before the droplet disappears. 

Increasing further the pressure above the kinetic $\alpha\to\beta$ transition
line, the free energy of the $\beta$ film becomes less than that of the $\alpha$
film {(Fig.5-\textbf{f})}. {Therefore, a droplet prepared on top of an $\alpha$ film
relaxes to a state where it stands on top of the preferred $\beta$ state.}
This corresponds to the `sunny side up' configuration found experimentally 
at sufficiently high saturation---see {Fig.~\ref{snapshots2}(\textbf{g-j})} and 
{Supplementary Movie~4};  
{\emph{c.f.}} Fig.~1-A from \cite{murata16}. 
Eventually the saturation is large enough that the $\beta$ film metastable
minimum is washed away by the linear term {$\Delta p_{\rm k}  h $} in
{$\omega_{\rm k}(h)$}. In
this case, the system becomes highly unstable (i.e.\ linearly unstable to
perturbations), and small satellite droplets can form, either in the
neighborhood of a larger droplet, or directly from a single local perturbation
on the solid surface ({Supplementary Figure~3} and {Supplementary
Movie~6}), a situation that very much resembles experimental observations -- see Movies S1 and S2 from Ref.~\cite{murata16}. Eventually, in the long time limit the inhomogeneities disappear completely, and the premelting film thickness diverges. Crystal growth then proceeds below a macroscopically thick wetting film that feeds on the surrounding bulk vapor.

\section{Discussion}

{In our study we have discussed ice premelting, but
our results rationalize the behavior of out-of-equilibrium premelting films and
wetting on reactive substrates quite generally. In particular, we see that for small deviations
away from the sublimation line, freezing occurs in a steady state regime with
constant film thickness. 
In this regime, the thickness of the premelting
film is dictated by an equilibrium interface potential and the underlying 
growth mechanism. For a given growth mechanism, our results show that the
outcome of the out of equilibrium dynamics may be predicted accurately from an underlying free energy
functional in analogy with  wetting  on inert substrates. As long as the system
remains in this steady state, the premelting film thickness is well defined and
depends both on temperature and pressure. 

We emphasize that it is not possible to interpret the dynamics of
the quasi liquid layer without taking into account the behavior of the underlying
substrate. In particular}, our results demonstrate that the complex dynamics of a buried solid surface can
be conveyed to the experimentally accessible outer surface of the quasi-liquid
film.
We also confirm that observation of terrace translation, spiral growth and
nucleation
observed in experiment is fully consistent with the existence of
a nanometer thick premelting film as observed in simulations
\cite{conde08,qiu18,benet16,llombart20b}. 
Accordingly, the motion of the experimentally accessible outer surface may
be used to interpret the hidden dynamics of the inner surface,
{very much in agreement with expectations of the Kuroda-Lacmann model
 \cite{kuroda82}.}

{The change from a thin to a thick film regime
   that occurs across well defined kinetic lines can result in a significant
   change in the mechanism for crystal growth. In the thin film regime,
   the growth of steps  is energetically expensive, because the nuclei are barely
   buried by the premelting film: steps formed feel a large inhomogeneity
   as the density changes from solid to vapor across a thin water film.
   As the kinetic coexistence line is
   traversed, however, liquid droplets condense on the ice surface.
   Steps formed below feel a much smaller tension, similar to that 
   at the ice/water interface. Their free energy of formation is
   therefore much smaller, and leads to a significant increase of the
   growth rate at places where droplets have condensed.
   This has immediate implications for our understanding of ice crystal
   growth \cite{nelson98,demange17}.  Since crystal
   corners have high local saturation, droplets are more likely to
   condense there, providing a source of water for the crystal to
   feed by a growth and spread mechanism from corners towards facet
   centers  as observed in experiments
   \cite{kobayashi67,nelson98,libbrecht19}. Furthermore, small crystallites with  large vapor pressure
   are more likely to have droplets condense at their corners, 
   explaining why the growth mechanism on a basal facet
   appears to be different in large and in small crystallites
   \cite{libbrecht19}. 
 Interestingly, this suggests that droplet
   condensation could  play a role in the tip splitting mechanism
   of ice grown from the vapor. Advanced  optical microscopy
   appears  a candidate technique for the verification of this hypothesis. 

In summary, we find a discontinuous change of crystal growth mechanisms with saturation.
Combined with recent findings of non-monotonic temperature dependence of step
free energies \cite{llombart20,llombart20b}, our results could help fill the gap between microscopic theories
and mesoscopic models of snow flake growth \cite{demange17}.
}

\section{Methods}

\subsection{Computer simulations}
Simulations of an equilibrated ice slab in the $NVT$ ensemble are performed in the
temperature range 210--270~K for the TIP4P/Ice model  \cite{abascal05} using 
GROMACS 5.0.5. {The equations of motion are integrated using
 the Leap-Frog algorithm, with a time step of 3~fs. Bond and angle constraints
 are applied using the LINCS algorithm. The canonical ensemble is sampled using
 thermostated dynamics with the velocity rescale algorithm \cite{bussi07}. The
 Lennard-Jones interactions are truncated at a distance of 9~{\AA}.
 Electrostatic interactions are evaluated using the Particle Mesh Ewald algorithm with the same
 real space cutoff. We calculate the reciprocal space term using a total of
 $80\times 64\times 160$  vectors in the $x$, $y$, $z$ reciprocal directions,
 respectively. We use a 0.1~nm grid spacing and fourth order interpolation
 scheme for the charge structure factor. 
 Simulations
 are carried out in systems consisting of $8\times8\times5$  unit cells of
 pseudo-orthorhombic geometry, each containing 16 molecules.
The initial
   configurations for the solid ice slab are prepared with a random realization
of the hydrogen bond network, following Ref.\cite{buch98}. One such initial
lattice is provided as {Supplementary Data 1}. This is then simulated at 1 bar
to obtain the equilibrium lattice parameters and placed in vacuum for further
equilibration in the $NVT$ ensemble during 15~ns. Averages are collected on
production runs 35~ns long.
During the  simulations, we identify structurally
liquid-like molecules using the $\bar{q}_6$ order parameter \cite{lechner08}.
Once these molecules are identified, we determine the locations of the
liquid-vapor and solid-liquid surfaces as explained in Ref.~\cite{benet19}.
From these two surfaces, we calculate the local film thickness as the difference
between these, {$ h (\mathbf{ x })=L_{\rm lv}(\mathbf{ x })-L_{\rm
sl}(\mathbf{ x })$}. For
the calculation of the interface potential, the local film thickness for a given
configuration is laterally averaged, in order to obtain the average liquid film
thickness. The set of global film thicknesses obtained 
are used to compute the probability histograms $P( h )$, from which
$g( h )$ can be calculated as detailed in the {Supplementary Note 1}. 
The results for $g(h)$ are fitted to the model described in the main
text. Parameter values and further details are given in {Supplementary
Table~1} and {Supplementary Note~3}.
}

\subsection{Gradient Dynamics}

{
Numerical computations of the dynamics of the thin-film equations are performed using the method of lines, similar to that used in Ref.~\cite{yin17}, but with a periodic pseudospectral method for the spatial derivatives. 
The method is extended to evolve the two interfaces (solid-liquid, and
liquid-vapor), with coupling terms involving mass transfer and the two interface
potentials naturally included. For the evolution of the solid-liquid
interface, a pinning effect in the horizontal direction can occur if too few
mesh points are used. Consequently, rather than using an extremely large number
of points in the finite difference scheme  we implement a periodic 
pseudospectral method which significantly increases the rate of numerical
convergence.  
The numerical method uses  discretization on
a regular (periodic) grid and a band-limited interpolant derived using the
discrete Fourier transform and its inverse to form the differentiation matrices
which act in real space. The presence of the 
premelting film avoids the need to explicitly evolve the 
contact lines, in comparison to some of our previous work using pseudospectral
discretisation \cite{DNS_Shikh,DNS_Crack}. For the time stepping, the ode15s
Matlab variable-step, variable-order solver is used. Our
numerical calculations are performed on the nondimensionalised version of the
model equations. We find that choosing $\kappa_1^{-1}\approx 0.49$~nm and
{ $\tau=3\eta\,(\kappa_1 \gamma_{\rm lv})^{-1}\approx 0.11$ }~ns as our units of length and time in
the nondimensionalisation works well.  Further details of the method and
initial conditions are given in the {Supplementary Note 8 and Supplementary
References}.
}

\subsection{Model Parameters}

Phase coexistence data required to compute {$\Delta p_{\rm sl}$},
{$\Delta p_{\rm lv}$},
structural properties of ice, and surface tension coefficients are obtained from
the literature as described in {Supplementary Table~2-3} and 
{Supplementary Note~6}. The kinetic growth coefficients
{$k_{\rm sl}$} is estimated from kinetic theory of gases, and
{$k_{\rm lv}$} is chosen such that the kinetic coexistence line has a slope
similar to experiments.  The sine Gordon coefficient 
{ $u=1.3\times 10^{-4}$ J m$^{-2}$ } is chosen to match step free energies
from the literature. The viscosity is taken from literature values of
undercooled water. Further details of the choice of model parameters 
are given in {Supplementary Note 7}. The actual model parameters
used in this work may be found in {Supplementary Tables 1-3}.

\section*{Acknowledgements}

We are indebted to Uwe Thiele for advice on formulating our gradient dynamics model. LGM is grateful to Loughborough University for hosting a stay funded by the `Programa Estatal de Promoci\'on del Talento y su Empleabilidad en I+D+i’ of the Spanish Ministerio de Educaci\'on, Cultura y Deporte (Plan Estatal de Investigaci\'on Cient\'ifica y T\'ecnica y de Innovaci\'on 2013–2016). We acknowledge the computer resources at MareNostrum and the technical support provided by BSC/MN  (QCM-2017-2-0008, QCM-2017-3-0034). PL, EGN and LGM were funded by the Spanish Agencia Estatal de Investigaci\'on under grant FIS2017-89361-C3-2-P and DNS by EPSRC grant EP/R006520/1.


\newpage

\onecolumngrid

\section*{How ice grows from premelting films and water droplets\\ (Supplementary Information)}

\section*{Supplementary Figure 1}

\begin{figure}[h!]
\centering
\includegraphics[width=0.5\columnwidth]{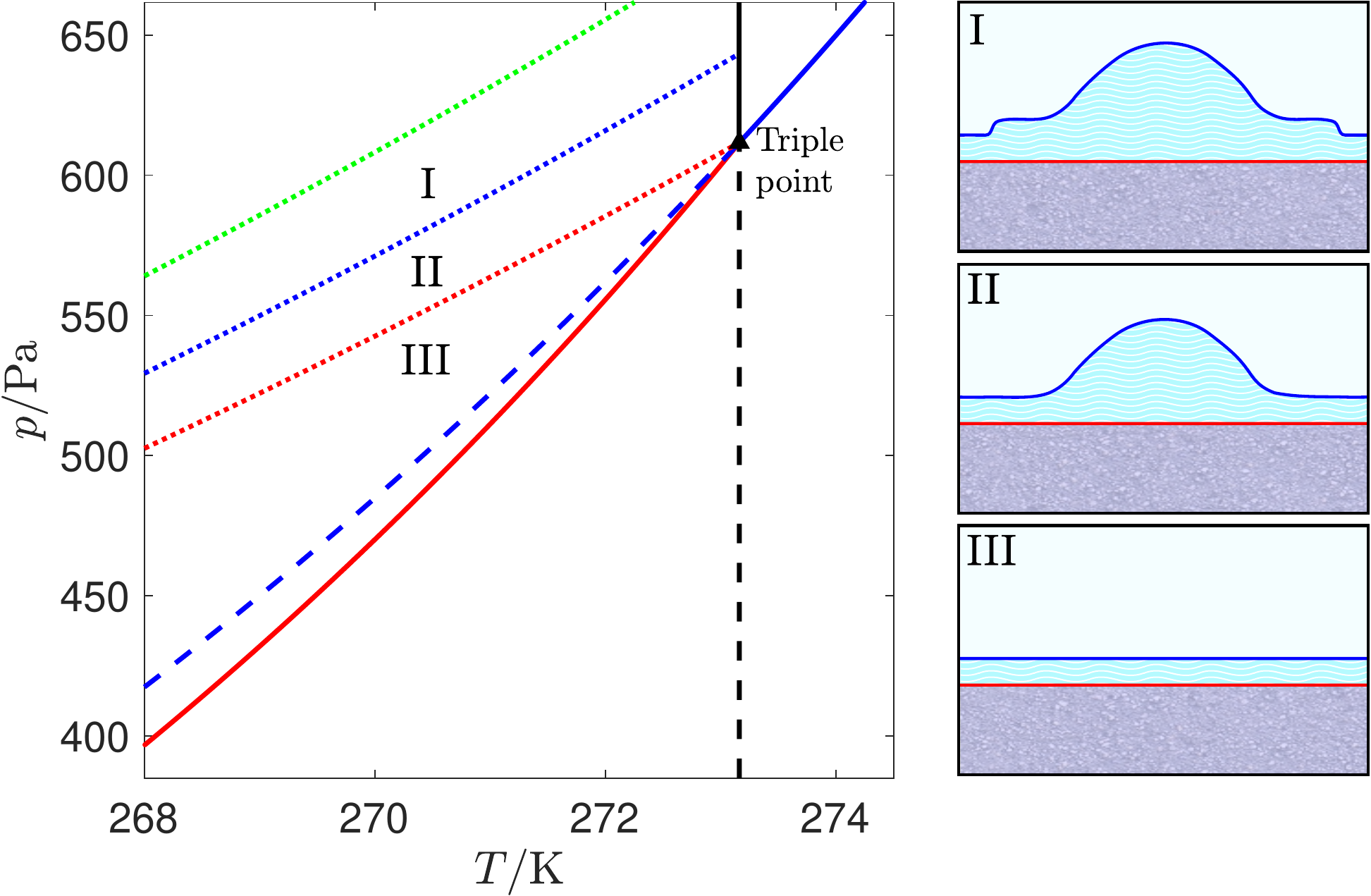}
\caption{
\label{paths}
   \footnotesize
         \textsf
{\bf Kinetic wetting phase diagram for the ice/vapor surface with
linear growth ($u=0$)}.
  On the left is the equilibrium phase diagram of water in the neighborhood of
  the triple point. The solid lines are
  the melting (black), vaporization (blue) and sublimation (red) lines. The
  dashed lines are metastable prolongations of the melting and vaporization
  lines. Dotted red and blue lines are {kinetic transition lines} which
	  describe the transitions observed in
  experiments between the states illustrated on the right, namely {(I) a
  spreading film below a droplet, (II) a droplet on top of a homogeneous surface and (III) a
  homogeneous surface} \cite{murata16}. The green dotted line is the kinetic spinodal line where quasi-stationary states are no longer 
  stable. The kinetic transition lines shown here 
  have been calculated using the model in {Eqs.~(3) and (4)} of the main text, assuming
  linear growth ($w=0$). {The  interface potential is scaled by a factor of 30 
in order to illustrate how the separation between kinetic
phase lines increases as the depth of the minima in the interface potential
increase (See {Fig.~3 of the main text with the associated discussion} and Supplementary Note 3 {below})}.
}
\end{figure}

\newpage

\section*{Supplementary Figure 2}

\begin{figure}[h!]
         \centering
	   \includegraphics[width=0.5\columnwidth]{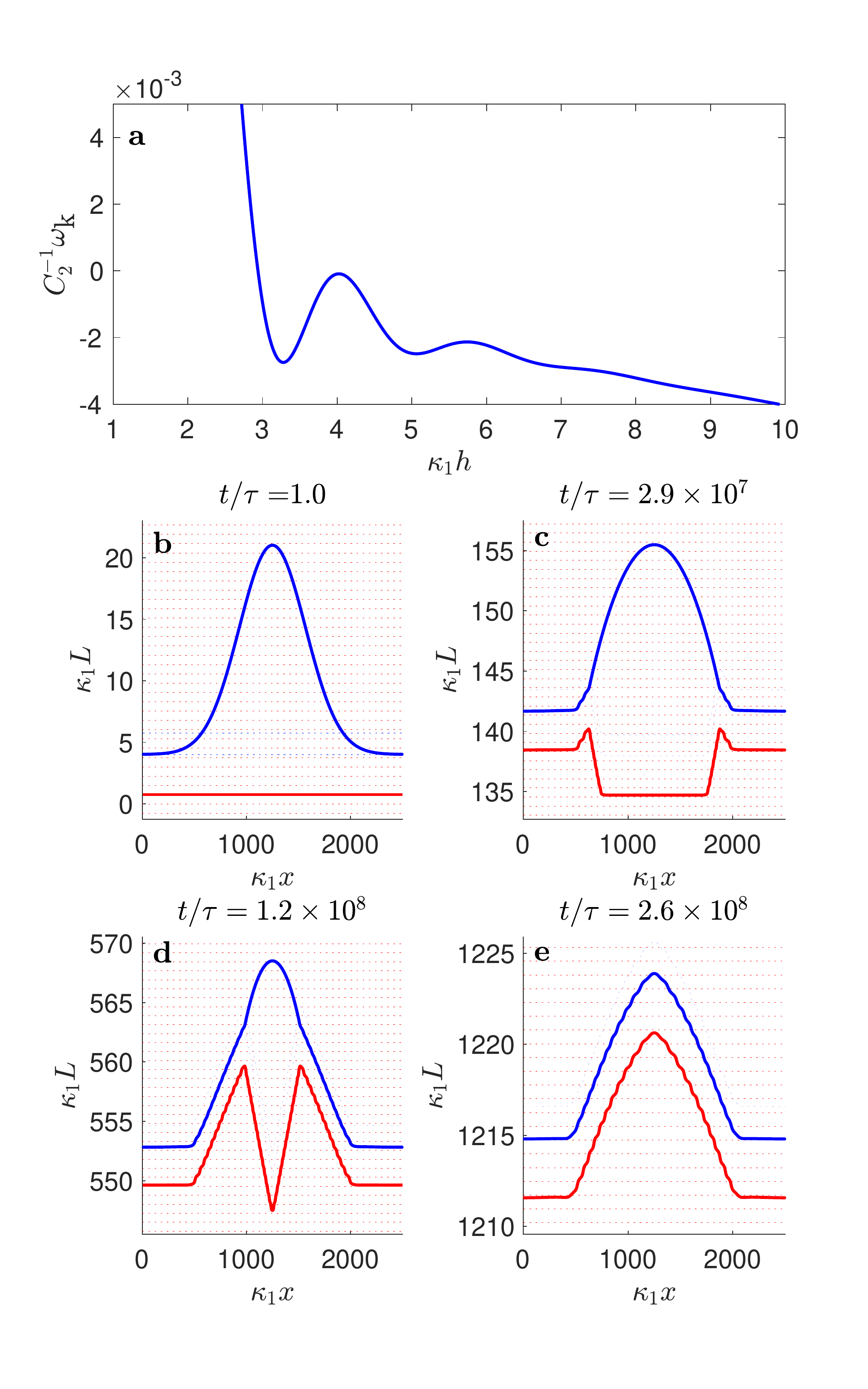}
        \caption{
                 \label{fig:s2}
		     {
			  {\bf Surface dynamics for an initial droplet 
			     at a state point above the
			  kinetic liquid-vapor coexistence line}, 
		     $(p,T)=$(517.5~Pa, 269.5~K),
		     corresponding to the {triangle ($\triangle$)} symbol in Fig.~3 {(main text)}.
		     {The color/style code of the lines is} as explained in captions to Fig.~4 {(main text)}.
		     Compared to Fig.~4 ({{\bf f}-{\bf j}}) {(main text)} for the dynamics of  the same
		     droplet right below the kinetic liquid-vapor coexistence,
		     the droplet is now stabilized for a long period, and a crater
		     is formed below.
		  }
           }
\end{figure}

\newpage

\section*{Supplementary Figure 3}

\begin{figure}[h!]
         \centering
	   \includegraphics[width=0.5\columnwidth]{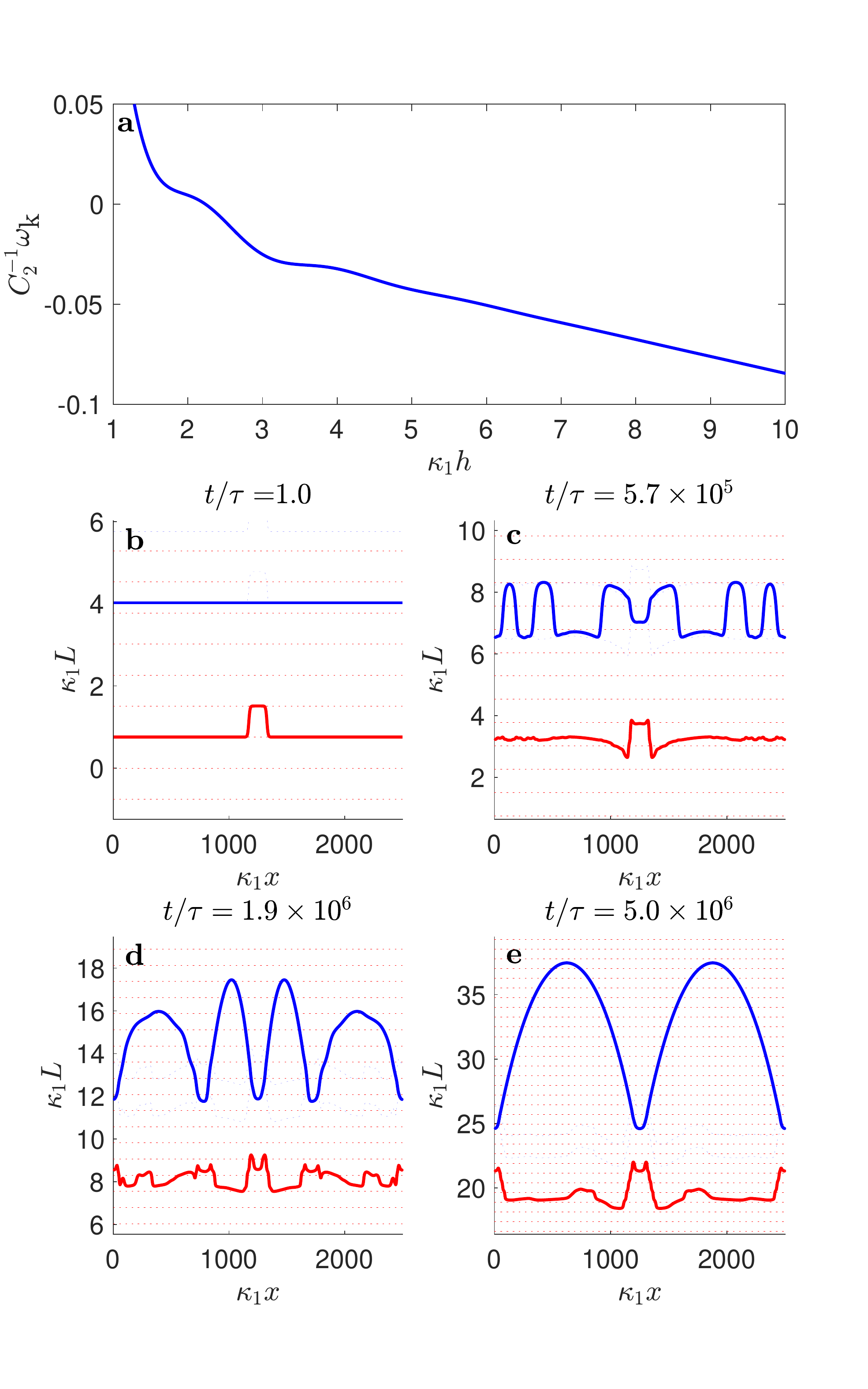}
        \caption{
                 \label{fig:s4}
		     {
			  {\bf Surface dynamics for a system with an initial terrace
			  at a state point above the kinetic spinodal line},
		     $(p,T)=$(535~Pa, 269.5~K), corresponding to the
		     {lozenge ($\lozenge$)}
		     symbol in Fig.~3 {(main text)}. {The color/style code of the lines is} as explained in 
		  captions to Fig.~4 {(main text)}.}
		     Compared to Fig.~4 ({{\bf b}-{\bf e}}) {(main text)} for the dynamics of  the same
		     initial terrace, the flat liquid-vapor surface becomes unstable and forms satellite droplets that grow and aggregate over time to leave the ice surface covered in a thick film of liquid.
           }
\end{figure}

\newpage

\section*{Supplementary Table 1}

\begin{table}[h!]
   \centering
   \begin{tabular}{cc}

	Property & Value \\
	\hline
	$C_1$ & $3.143\times 10^{-3}$ \textrm{J m}$^{-2}$\\
	$C_2$ & $4.116\times 10^{-2}$ \textrm{J m}$^{-2}$\\
	$\kappa_1$ & $2.043\times 10^9$ \textrm{m}$^{-1}$\\
	$q_0$ & $7.148\times 10^9$ \textrm{m}$^{-1}$ \\
	$\alpha$ & $5.144$ \\
	$B$ & $7.875\times 10^{-31}$ \textrm{Jm}\\
	$f$ & $1.106$ \textrm{(unitless)}\\
	$a$ & $3.03\times 10^7$ \textrm{m}$^{-1}$\\
	$b$ & $5.0\times 10^8$ \textrm{m}$^{-1}$\\
	\hline
	$\alpha$-minimum & $1.6$~nm \\
	$\beta$-minimum & $2.4$~nm \\
	$\Pi^*$, $\alpha-\beta$ transition & $-4.60\times 10^4$~Pa \\
	$\Pi^*$, $\beta$ spinodal & $-1.02\times 10^5$~Pa \\
	\hline
   \end{tabular}
   \caption{
	\label{tab:g_consts}
	{\bf Parameters used in the interface potential}, $g(h)$
	with details on the locations of the $\alpha$ and $\beta$ minima
	and spinodals. Further details of the fitting procedure may be found
	in {Supplementary Note 3 below}.
   }
\end{table}

\section*{Supplementary Table 2}

 \begin{table*}[h!]
   \centering
    \label{tpd}
    \begin{tabular}{ccc}
    \hline
	 Property & Value & Reference \\
	 \hline
	 {$T_\textrm{t}$} & 273.16 K &  \cite{wagner02} \\
	 {$T_\textrm{t}$} & 0.1 $^\circ$C &  \cite{wagner02} \\
	 {$p_\textrm{t}$} & 611.65 Pa & \cite{wagner02}   \\
	 {$\rho\liq$} & 55 498  mol m$^{-3}$ & \cite{wagner02} \\
	 {$\rho\sol$} & 50 888  mol m$^{-3}$ & \cite{feistel06} \\
	 {$\rho\vap$} & 0.2694  mol m$^{-3}$ & \cite{wagner02} \\
	 {$\Delta H{\sv}$} & 51 059 J mol$^{-1}$ & \cite{murphy05} \\
	 {$\Delta H{\lv}$} & 45 051 J mol$^{-1}$ & \cite{murphy05} \\
	 {$\Delta H{\sl}$} & 6 008 J mol$^{-1}$   & \cite{murphy05} \\
	 \hline
    \end{tabular}
   \caption{   {\bf Triple point data of water}. These results
	are used for the calculation of
    thermodynamic functions described in the Supplementary Note 6. 
 Conversion from mass to molar units performed assuming {$M_\textrm{w}=18.015$}~g/mol.}
 \end{table*}

\newpage

\section*{Supplementary Table 3}

\begin{table}[h!]
\centering
\begin{tabular}{ccc}
Property & Value & Source \\ 
\hline
{$d_\textrm{B}$} & $0.37\times 10^{-9}$ m &    \cite{pruppacher10} \\
{$\rho_\textrm{lv}$} & {$p_\textrm{lv}/(R_\textrm{c} T)$} & ideal gas law \\
{$T_\textrm{c}$} & $T-273.15$ $^\circ$C & Celsius scale \\
{$\rho_\textrm{l}$} & \begin{tabular}{@{}c@{}}{$55502 + 3.4549T_\textrm{c} - 0.44461T_\textrm{c}^2\ldots$} \\
{$\ldots + 0.0028885T_\textrm{c}^3 - 0.00031898T_\textrm{c}^4$} mol m$^{-3}$\end{tabular} &
\cite{hare87,tanaka01}
 \\
 {$\rho_\textrm{s}$} & {$50885 - 9.71T_\textrm{c} - 0.03T_\textrm{c}^2$} mol m$^{-3}$ &   \cite{pruppacher10} \\
 {$\gamma_\textrm{sl}$} & {$(28 + 0.25 T_\textrm{c})\times 10^{-3}$} J m$^{-2}$ & \cite{pruppacher10} \\
 {$\gamma_\textrm{lv}$} & {$(75.7 - 0.1775 T_\textrm{c})\times 10^{-3}$} J m$^{-2}$ & \cite{fletcher70} \\
$\eta$ & $1.39\times 10^{-4}(T/225 - 1)^{-1.64}$ kg m$^{-1}$ s$^{-1}$ &
\cite{taborek86} \\
$u$ & $1.3\times 10^{-4}$ J m$^{-2}$ & This work \\
{$k_\textrm{lv}$} & {$3.4\times 10^{-10}\rho_\textrm{lv}T^{-1/2}\times 10^{-3}$} m s$^{-1}$
Pa$^{-1}$  &  Knudsen-Hertz law \\
{$k_\textrm{sl}$} & {$6.4k_\textrm{lv}$} & Slope of phase line \\
\hline
\end{tabular} 
\label{tab:mod_consts}
\caption{
{\bf
Temperature dependent coefficients for use in the mesoscopic calculations}. 
$T$ refers to absolute temperature in K. $T_c$ refers to temperature in the
Celsius scale. Further details on the derivation of these coefficients may
be found in {Supplementary Note 7 below}.}
\end{table}

\section*{Supplementary Note 1: Numerical calculation of the interface potential}

\subsection*{Definition of the interface potential}

The excess grand potential $\Omega$ (Landau free energy) per unit area for a liquid film of thickness $\hh$ on a planar solid surface in equilibrium with a bulk vapor phase with chemical potential $\mu$ and temperature $T$ is
{\begin{eqnarray}\label{eq:ghd}
   \frac{\Omega+p_\textrm{v}V}{A} &=& \gamma_\textrm{sl}+\gamma_\textrm{lv}+g(\hh;T) - \Delta p\lv(T,\mu)\hh\\
   &\equiv& \gamma_\textrm{sl}+\gamma_\textrm{lv}+\omega(\hh;T,\mu),
\end{eqnarray} }
where $V$ is the volume of the system, $A$ is the area of the surface, {$\gamma_\textrm{sl}$} is the solid/liquid interfacial tension, {$\gamma_\textrm{lv}$} is the liquid/vapor interfacial tension, $g(\hh)$ is the interface potential for the film at liquid-vapor coexistence, often referred to as the binding potential, and {$\Delta p\lv(T,\mu)= p_\textrm{l}(T,\mu) - p_\textrm{v}(T,\mu)$} is the pressure difference of the bulk liquid and vapor phases at the chemical potential of the bulk vapor. The potential $\omega(\hh;T,\mu)$ is the effective interface potential that determines the interfacial phase behavior.

In relevant previous work, the interface potential of liquid films adsorbed on an inert substrate was calculated by performing grand-canonical simulations at liquid-vapor coexistence \cite{macdowell05,grzelak08}. In that case, $\omega(\hh;T,\mu) = g(\hh;T)$, and the free energy may be evaluated from {$A \, \omega(\hh;T,\mu) = - k_\textrm{B}T\ln P(\hh)$}, where $P(\hh)$ is the probability distribution of $\hh$, collected during the grand canonical simulation with enhanced sampling techniques and {$k_\textrm{B}$} is Boltzmann's constant. Effectively, the procedure is equivalent to performing a series of canonical simulations at different film thicknesses \cite{benet14b}.

For the case of a one component system with liquid adsorbed at the solid/vapor interface, the above method cannot be applied, because the three phase system at fixed temperature only exists at equilibrium at the solid/vapor coexistence chemical potential. Instead, we perform a set of fixed-$NVT$ simulations at different temperatures ($N$ is the number of molecules), similar to previous calculations in studies of the interface potential for grain boundary premelting \cite{hoyt09,hickman16}.

For a liquid film adsorbed at the solid/vapor interface along the sublimation line ({$T,\mu_\textrm{sv}(T)$}), {Supplementary} \Eq{ghd} gives
{\begin{equation}
   \omega(\hh;T,\mu\sv) = g(\hh;T) - \Delta p\lv(T)|\sv \hh,
\end{equation} }
where $g(\hh;T)$ is the interface free energy for the film along the liquid-vapor coexistence line, and {$\Delta p\lv(T)|\sv  = p_\textrm{l}(T,\mu\sv)-p_\textrm{v}(T,\mu\sv) $} is the pressure difference between liquid and vapor bulk phases at the solid-vapor coexistence chemical potential.

\begin{figure}[t]
         \centering
	   \includegraphics[width=.45\linewidth]{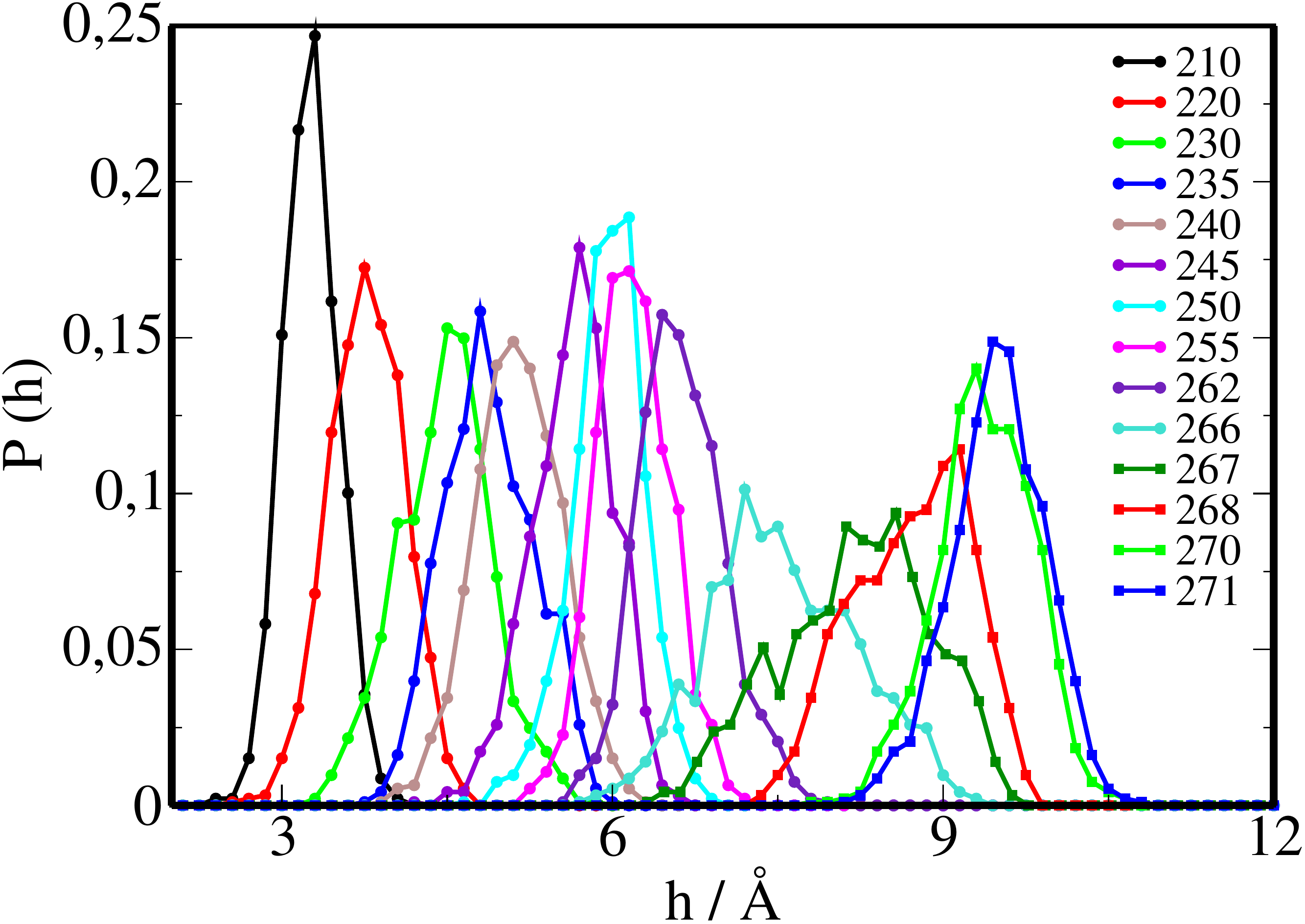}
	   \includegraphics[width=.45\linewidth]{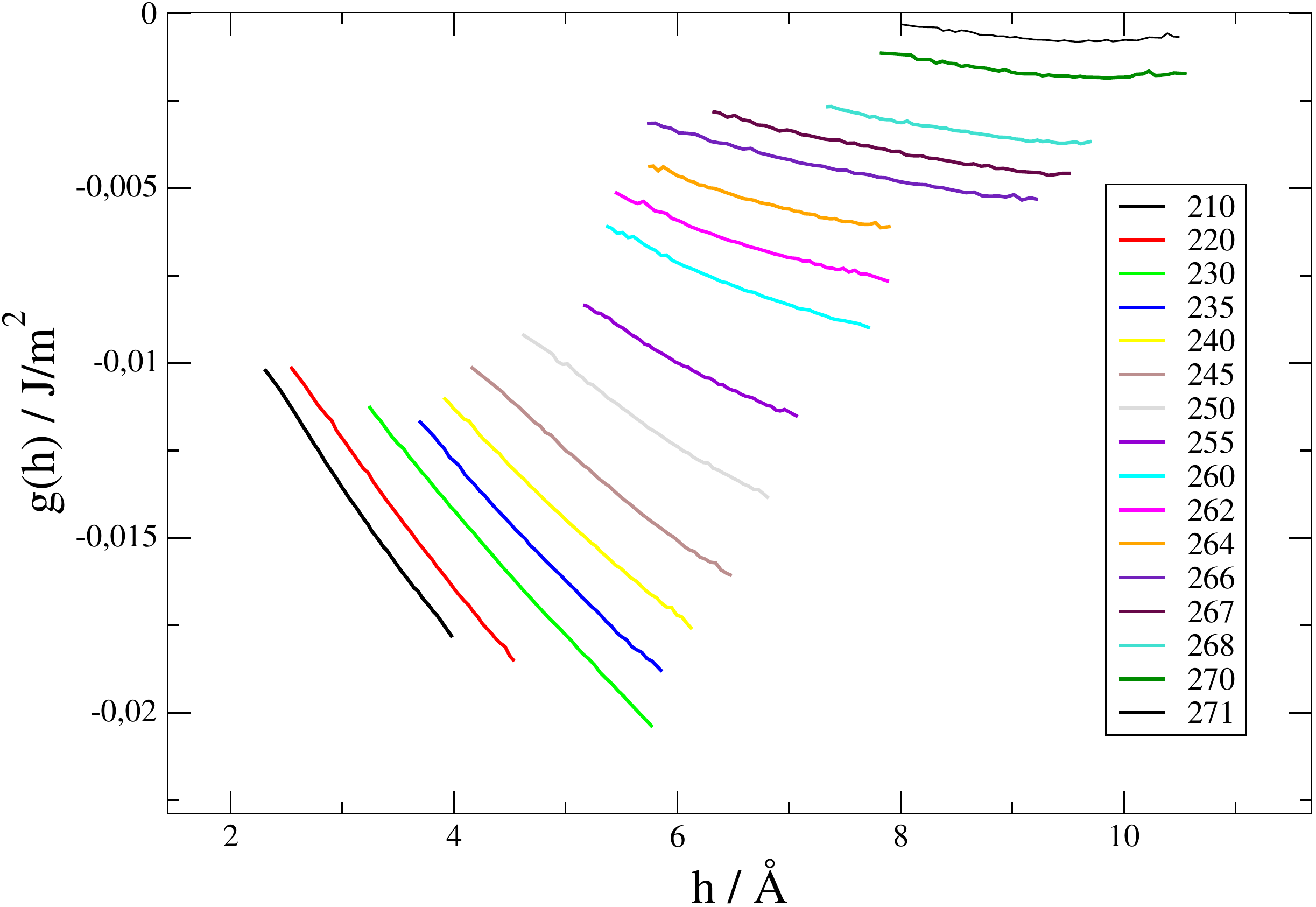}
        \caption{
            \label{fig:ghcalc}
		Left: The global film height probability distribution, obtained from a sequence of independent simulations at fixed $NVT$ and for a range of different temperatures (210-271 K), as given in the key. Right: The corresponding piecewise interface potentials.}
   \end{figure}

Performing simulations of the solid phase at constant temperature, initiated in a vacuum, the system equilibrates into a state of solid/vapor coexistence, with a premelting liquid film at the interface with thickness dictated by imposed thermodynamic conditions. At this temperature, the film thickness fluctuates according to a probability distribution {$P(\hh;T,\mu\sv)$}, which can easily be collected during the course of the simulation, as shown in {Supplementary} Fig.~\ref{fig:ghcalc}.

The interface potential in the range of observed film thicknesses may be calculated as
{\begin{equation}\label{eq:gcalc}
   g(\hh;T) = -\frac{k_BT}{A}\ln P(\hh;T,\mu\sv) +  \Delta p\lv(T)|\sv \hh + C_T,
\end{equation} }
where $C_T$ is an arbitrary constant. By performing a sequence of simulations at different temperatures, one obtains a set of piecewise potentials $g(\hh;T_i)$, which overlap for small ranges of $\hh$, provided the simulations are performed at sufficiently close temperature intervals. The right hand panel of {Supplementary} Fig.~\ref{fig:ghcalc} shows the set of piecewise functions obtained at a series of different temperatures, with values as indicated in the key. Since the temperature dependence of $g(\hh;T)$ is small, the piecewise function can be combined into a single continuous interface potential by choosing suitable constants $C_{T_i}$. The resulting function is continuous and shows no apparent singularities, consistent with the assumption of weak temperature dependence of the various piecewise terms $g(\hh;T_i)$.

\subsection*{Calculation of the pressure difference {$\Delta p\lv(T)|\sv$}}

In order to evaluate the interface potential, we must first determine {$\Delta p\lv(T)|\sv$}. We start from the Gibbs-Duhem thermodynamic relation
\begin{equation}\label{eq:Gibbs_Duhem}
  Nd\mu = - S dT + V dp,
\end{equation}
where $S$ is the entropy. From this we obtain the following equivalent pair of relations
\begin{equation}\label{eq:dmu}
  d\mu = - s dT + \frac{1}{\rho} dp,
\end{equation}
\begin{equation}\label{eq:dp}
  d p = \rho s dT + \rho d\mu,
\end{equation} 
where $s=S/N$ is the entropy per particle and $\rho=N/V$ is the number density. At phase coexistence, $\mu$, $p$ and $T$ are equal in the two coexisting phases. Hence, along the the solid (subscript {$\textrm{s}$}) and vapor ({$\textrm{v}$}) coexistence line we have {$d\mu_\textrm{s}=d\mu_\textrm{v}$, $dp_\textrm{s}=dp_\textrm{v}$ and $dT_\textrm{s}=dT_\textrm{v}$}. Therefore, from the first of these together with {Supplementary} \Eq{dmu} we obtain the familiar Clausius-Clapeyron equation for the variation of the vapor pressure along the sublimation line
{\begin{equation}\label{eq:clausiusp}
   \left .   \frac{d p}{d T} \right |\sv = \rho_s\rho\vap \frac{ s_\textrm{s}-s_\textrm{v}
   }{\rho\vap - \rho\sol }.
\end{equation} }
Similarly, from {Supplementary} \Eq{dp} we obtain
{\begin{equation}\label{eq:clausiusm}
   \left .   \frac{d \mu}{d T} \right |\sv = \frac{\rho\vap
   s\vap - \rho\sol s\sol}{\rho\sol - \rho\vap }.
\end{equation}}
Thus, from {Supplementary} \Eq{clausiusp} the variation of vapor pressure along the sublimation line is
{\begin{equation}\label{eq:pvap}
   d p\vap |\sv = \rho_s\rho\vap \frac{ s_\textrm{s}-s_\textrm{v} }{\rho\vap - \rho\sol } dT,
\end{equation} }
whereas the pressure variations of the liquid phase is given more generally by {Supplementary} \Eq{dp} as
{\begin{equation}
  d p\liq = \rho\liq s\liq dT + \rho\liq d\mu.
\end{equation} }
However, we must evaluate the liquid pressure along the sublimation line, so $\mu$ is not an independent variable.  Rather, it is given by the Clausius-Clapeyron type {Supplementary} \Eq{clausiusm}, and thus
{\begin{equation}\label{eq:pliq}
  d p\liq|\sv = \rho\liq s\liq dT + 
   \rho\liq \frac{\rho\vap s\vap - \rho\sol s\sol}{\rho\sol - \rho\vap } dT.
\end{equation} }
Therefore, the variation of {$d (p\liq - p\vap)|\sv$} along the sublimation line is obtained from {Supplementary} Eqs.~(\ref{eq:pvap}) and (\ref{eq:pliq}) after some rearrangements, as
{\begin{equation}\label{eq:long_dp}
   d (p\liq - p\vap)|\sv = \left . \frac{
\rho_\textrm{s}\rho_\textrm{l}s_\textrm{l}-
\rho_\textrm{v}\rho_\textrm{l}s_\textrm{l}+
\rho_\textrm{l}\rho_\textrm{v}s_\textrm{v}-
\rho_\textrm{l}\rho_\textrm{s}s_\textrm{s}+
\rho_\textrm{s}\rho_\textrm{v}s_\textrm{s}-
\rho_\textrm{s}\rho_\textrm{v}s_\textrm{v}}{\rho\sol - \rho\vap } \right |\sv dT.
\end{equation} }
In principle, this equation could be integrated starting from the triple point, where {$p\liq - p\vap=0$}, down to lower temperatures, by using experimental or simulation data for entropies and densities along the sublimation line.  A zeroth order integrated form of this equation  may be found in Elbaum and Schick \cite{elbaum91b}.

Here, we take a different more convenient approach by expressing this equation in terms of liquid-vapor and solid-vapor coexistence pressures, which are known from experiments with great accuracy. To achieve this, we first notice {$\rho\vap\ll\rho\liq$} and {$\rho\vap\ll\rho\sol$}. Therefore, the exact result in {Supplementary} \Eq{long_dp} can be greatly simplified with only a very small loss in accuracy, to
{\begin{equation}\label{eq:dplv}
        d (p\liq - p\vap)|\sv = - \rho\liq (s_\textrm{s}-s_\textrm{l}) dT|\sv.
\end{equation} }
Now, we write
{\begin{equation}\label{eq:s_shuffle}
     (s_\textrm{s}-s_\textrm{l})|\sv  = \left [ (s\sol - s\vap) - (s\liq - s\vap) \right ]\sv.
\end{equation} }
Furthermore, assuming the vapor behaves as an ideal gas so that (i) {$s=-k_\textrm{B}(\ln(\Lambda^3\rho)-1)$}, where $\Lambda$ is the thermal de Broglie wavelength, and (ii) {$\rho=p/k_\textrm{B}T$}, we can write the vapor entropy at the sublimation line in terms of the vapor entropy at the condensation line as
{\begin{equation}\label{eq:s_v_ideal}
   s\vap|\sv = s\vap|\lv + k_\textrm{B}\ln \frac{p\lv}{p\sv}.
\end{equation} }
Substituting this into {Supplementary} \Eq{s_shuffle} and noting that the entropy of the incompressible liquid phase hardly changes at all, which means that we may approximate {$s\liq|\sv = s\liq |\lv$}, so that from {Supplementary} \Eq{s_shuffle} and {Supplementary} \Eq{s_v_ideal} we can write
{\begin{equation}
   (s_\textrm{s}-s_\textrm{l})|\sv  = (s_\textrm{s}-s_\textrm{v})|\sv  - (s_\textrm{l} - s_\textrm{v})|\lv+  k_\textrm{B}\ln
   \frac{p\lv}{p\sv}.
\end{equation} }
Substituting this into {Supplementary} \Eq{dplv} then yields:
{\begin{equation}\label{eq:17}
   d (p\liq - p\vap)|\sv = -  \rho\liq \left [ (s_\textrm{s}-s_\textrm{v})|\sv  - (s_\textrm{l} - s_\textrm{v})|\lv+  k_\textrm{B}\ln
   \frac{p\lv}{p\sv} \right ] dT,
\end{equation} }
where now both {$(s_\textrm{s}-s_\textrm{v})|\sv$} and {$(s_\textrm{l} - s_\textrm{v})|\lv$} are actual entropies of
phase change. Invoking the Clausius-Clapeyron {Supplementary} \Eq{clausiusp} for these two quantities, assuming {$\rho_\textrm{v}\ll\rho_\textrm{s}$}, {$\rho_\textrm{v}\ll\rho_\textrm{l}$} and making the ideal gas approximation {$p=k_\textrm{B}T\rho$}, we obtain
{\begin{equation}
 \left . - (s_\textrm{s}-s_\textrm{v})|\sv = \frac{k_\textrm{B}T}{p\sv}\frac{d p}{dT}\right|\sv,
\end{equation} }
and a similar expression for {$(s_\textrm{l} - s_\textrm{v})|\lv$}. Substituting these into {Supplementary} \Eq{17}, we obtain the sought expression for {$d (p\liq - p\vap)|\sv$} explicitly in terms of vapor pressures along sublimation and condensation lines as
{\begin{equation}
   d (p\liq - p\vap)|\sv = \rho\liq d\left( k_\textrm{B}T \ln \frac{p\sv}{p\lv} \right ).
\end{equation} }
Integrating this equation from the triple point to a desired arbitrary temperature, we obtain
{\begin{equation}
   \Delta p\lv(T)|\sv =  \rho\liq k_\textrm{B}T \ln \frac{p\sv}{p\lv}.
\end{equation} }
This is the same result obtained in \cite{llombart20} by alternative means. We use explicit expressions obtained for the vapor pressures of the TIP4P/Ice model to calculate the required pressure difference for use in {Supplementary} \Eq{gcalc}.

\section*{Supplementary Note 2: Analytical formula for the surface van der Waals forces}

Elbaum and Schick calculated the van der Waals force contributions to the interface potential using Lifshitz theory \cite{elbaum91b}. The results are obtained only in numerical form from quadrature, which is not convenient for numerical purposes. Here we derive an accurate analytical approximation, along the lines of Ref.~\cite{macdowell19}.

Quite generally, the van der Waals forces between two media, $1$ and $2$, across a media $m$ enclosed between infinite slabs of media $1$ and $2$, give rise to an interface potential of the form
{\begin{equation}\label{eq:gh}
    g_\textrm{vdw}(\hh) = - \frac{A(\hh)}{12\pi\hh^2},
\end{equation} }
where $A(\hh)$ is the Hamaker function. In a well known approximation to Lifshitz theory, this is given as
{\begin{equation}\label{eq:hamakerf}
   A(\hh) = \frac{3}{2} k_\textrm{B} T \sum_{n=0}^{\infty \prime} 
            R(\omega_n) [ 1 + r_n ] e^{-r_n},
\end{equation} }
where the prime indicates that the first term is weighted by a factor of $1/2$, $r_n=2\epsilon_m^{1/2}\omega_n\hh/c$,  $\omega_n=\omega_T n$, $\omega_T=2\pi k_BT/\hbar$, and $\epsilon_m$ is the dielectric constant of the layer of thickness $\hh$. The function  $R(\omega_n)$ is a complicated expression that depends on the frequency dependent dielectric constants of the material and the film thickness $\hh$ \cite{macdowell19}. For practical purposes, it can be approximated via the simpler expression
\begin{equation}
   R(\omega_n) = \left (  \frac{\epsilon_1 - \epsilon_m}{\epsilon_1 + \epsilon_m} \right )
    \left (  \frac{\epsilon_2 - \epsilon_m}{\epsilon_2 + \epsilon_m} \right ),
\end{equation} 
where $\epsilon_1$ and $\epsilon_2$ are the frequency dependent dielectric constants of the media enclosing the layer of thickness $\hh$. At this stage it is convenient to single out the $n=0$ term in  {Supplementary} 
\Eq{hamakerf}, and to further approximate the remaining sum into an integral. Then
{\begin{equation}
g_\textrm{vdw}(\hh) = - \frac{A_{\omega=0}}{12\pi\hh^2} -  \frac{A_{\omega>0}(\hh)}{12\pi\hh^2},
\end{equation}}
where
{\begin{equation}
   A_{\omega=0}  = \frac{3}{4} \left (  \frac{\epsilon_1 - \epsilon_m}{\epsilon_1 +
   \epsilon_m} \right )
       \left (  \frac{\epsilon_2 - \epsilon_m}{\epsilon_2 + \epsilon_m} \right )
	 k_\textrm{B} T,
\end{equation} }
and
\begin{equation}\label{eq:Ahh}
   A_{\omega>0}(\hh) = \frac{3\hbar c}{8\pi\epsilon_m^{1/2}}   \int_{\nu_T}^{\infty} R(\nu) [ 1 + \nu \hh ] e^{-\nu\hh} d\nu,
\end{equation} 
where the sum over angular frequencies has been transformed into an integral over wavenumbers $\nu=2\epsilon_m^{1/2}\omega/c$ and $\nu_T=2\epsilon_m^{1/2}\omega_T/c$.

Elbaum and Schick parametrized the dielectric properties of water and ice, and argued that the term $(\epsilon_i - \epsilon_w)$ of the function $R(\nu)$ changes sign at ultra-violet frequencies, such that $R(\nu)<0$ in the infra-red, but $R>0$ at the extreme ultra-violet and beyond. In view of this, we split the integral of {Supplementary} \Eq{Ahh} and write:
{\begin{equation}
   A_{\omega>0}(\hh)  =  \frac{3\hbar c}{8\pi\epsilon_m^{1/2}} \int_{\nu_T}^{\nu_\textrm{UV}} R(\nu) [
	     1 + \nu \hh ] e^{-\nu\hh} d\nu + 
	     \frac{3\hbar c}{8\pi\epsilon_m^{1/2}} \int_{\nu_\textrm{UV}}^{\infty} R(\nu) [
		             1 + \nu \hh ] e^{-\nu\hh} d\nu,
\end{equation} }
where {$\nu_\textrm{UV}$} is the frequency at which $R(\nu)$ is maximum. The first integral can now be evaluated  using the first mean value theorem, and the second using the second mean value theorem, yielding
{\begin{equation}
   \begin{array}{ccc}
   A_{\omega>0}(\hh) & =  & \frac{3\hbar c}{8\pi\epsilon_m^{1/2}\hh} R(\nu_\textrm{IR})
[ (2 + \nu_T\hh) e^{-\nu_T\hh}  - ( 2 + \nu_\textrm{UV} \hh ) e^{-\nu_\textrm{UV}\hh} ]  \\ &
& + 
	    \frac{3\hbar c}{8\pi\epsilon_m^{1/2}\hh} R(\nu_\textrm{UV}) [ (2 + \nu_\textrm{UV}\hh)
	    e^{-\nu_\textrm{UV}\hh} - ( 2 + \nu_{\infty} \hh ) e^{-\nu_{\infty}\hh} ].
   \end{array}
\end{equation} }
This is an exact quadrature for  suitably chosen frequencies {$\nu_\textrm{IR}$} and $\nu_{\infty}$, satisfying {$\nu_T< \nu_\textrm{IR} < \nu_\textrm{UV}$}, and {$\nu_\textrm{UV}<\nu_{\infty}< \infty$}. Collecting terms, the above expression simplifies to
{\begin{equation}\label{eq:DLPM}
   A_{\omega>0}(\hh) =   \frac{3\hbar c}{8\pi\epsilon_m^{1/2}\hh} R(\nu_\textrm{IR})  \left [
	     (2 + \nu_{T} \hh ) e^{-\nu_T \hh}
	     + (f-1) (2 + \nu_\textrm{UV} \hh) e^{-\nu_\textrm{UV} \hh}
           - f (2 + \nu_{\infty} \hh) e^{-\nu_{\infty} \hh}
	    \right ],
 \end{equation}}
where {$f=R(\nu_\textrm{UV})/R(\nu_\textrm{IR})$}. {Supplementary} \Eq{DLPM} provides a simple analytic expression which properly captures the crossover from retarded to non retarded interactions, as well as the suppression of retarded interactions at large distances and the temperature dependence of the van der Waals forces.

Assuming that  the relevant wave-numbers are well separated, such that {$\nu_T \ll \nu_\textrm{UV} \ll \nu_{\infty}$}, we find the following four distinct regimes as  $\hh$ increases:
 \begin{itemize}
  \item The subnanometer range, $\nu_{\infty}\hh \ll 1$, describes either
	 the $\hh\to 0$ or $T\to 0$ behavior of $A_{\omega>0}$. Expanding
	 all the exponentials in {Supplementary} \Eq{DLPM}, one finds that
	 the terms of order $\hh^0$ inside the square brackets cancel
	 exactly. Retaining then the leading order terms in $\hh$, one finds
	 {\begin{equation}
	    A_{\omega>0}(\hh) =   \frac{3\hbar \omega_{\infty}}{4\pi}R(\nu_\textrm{UV}).
	    \end{equation}}
	    In this regime $ A_{\omega>0}$ recovers the standard low
	    temperature
	    asymptotic limit that is well known in the literature. In particularly,
	    $ A_{\omega>0}$  is independent of $\hh$  and one can talk
	    appropriately of a Hamaker constant. 
	 \item For {$\nu_\textrm{UV}\hh \ll 1 \ll \nu_{\infty}\hh$}, the last
	    term in {Supplementary} \Eq{DLPM} is exponentially suppressed, and  $ A_{\omega>0}$
	    develops  an explicit $\hh$ dependence
	   { \begin{equation}
		 A_{\omega>0}(\hh) = \frac{3\hbar c}{4\pi\epsilon_m^{1/2} \hh} R(\nu_\textrm{UV}).
		 \end{equation}}
		 Using this expression in {Supplementary} \Eq{gh}, we recover the standard
		 result
		 for retarded van der Waals interactions. In this range, the
		 free energy has naturally shifted from an $\hh^{-2}$ to an 
		 $\hh^{-3}$
		 dependence, while  the sign of the interactions remains
		 dominated by the UV dielectric response.
	    \item For {$\nu_{T}\hh \ll 1 \ll  \nu_\textrm{UV}\hh$}, the last
		 two terms of {Supplementary} \Eq{DLPM} are suppressed, and the retarded
		 interactions cross over from an ultraviolet dominated regime,
		 to an infrared dominated regime
	{\begin{equation}
	   A_{\omega>0}(\hh) = \frac{3\hbar c}{4\pi \epsilon_m^{1/2} \hh} R(\nu_\textrm{IR}),
	   \end{equation}}
	   since {$R(\nu_\textrm{IR})$} and {$R(\nu_\textrm{UV})$} have opposite signs, the
	   Hamaker function changes sign from positive to negative
	   as the film thickness becomes larger than the cross-over
	   wave-length {$\nu_\textrm{UV}$} lying
	   in the nanometer length scale.
	\item Finally, for $\nu_{T}\hh \gg 1$,  only the first term of {Supplementary} \Eq{DLPM} remains.
	   This results in  an exponentially decaying retarded interaction
	   corresponding to the expected suppression of $A_{\nu>0}$ at microwave
	   distances \cite{parsegian70,parsegian06}, with
	   {\begin{equation}
		A_{\omega>0}(\hh) = 3 k_BT R(\nu_\textrm{IR}) e^{-\nu_T \hh}.
	 \end{equation}}
 \end{itemize}

For practical purposes, we are only interested in modeling van der Waals forces out to distances of the order of decades of nanometers from the surface, so we assume $\nu_T\hh\ll 1$, and simplify {Supplementary} \Eq{DLPM} to
 {\begin{equation}\label{eq:crossover2}
      g_\textrm{vdw}(\hh)= -\frac{B}{\hh^3}\left [
	   1 - f \exp(-\nu_\textrm{UV} \hh) - (1-f) \exp(-\nu_{\infty} \hh)
			  \right ],
 \end{equation} }
where now  $B$, $f$, {$\nu_\textrm{UV}$} and $\nu_{\infty}$ are  parameters chosen to best model the results of Elbaum and Schick in the range of 1 to 10 nm. For sufficiently large $f>1$, this equation gives the expected crossover in the decay form of $g(h)$ from $\sim h^{-2}$ to $\sim h^{-3}$ dominated regimes found for the ice/water/air interface.

\begin{figure}[t]
         \centering
	   \includegraphics[width=.5\linewidth]{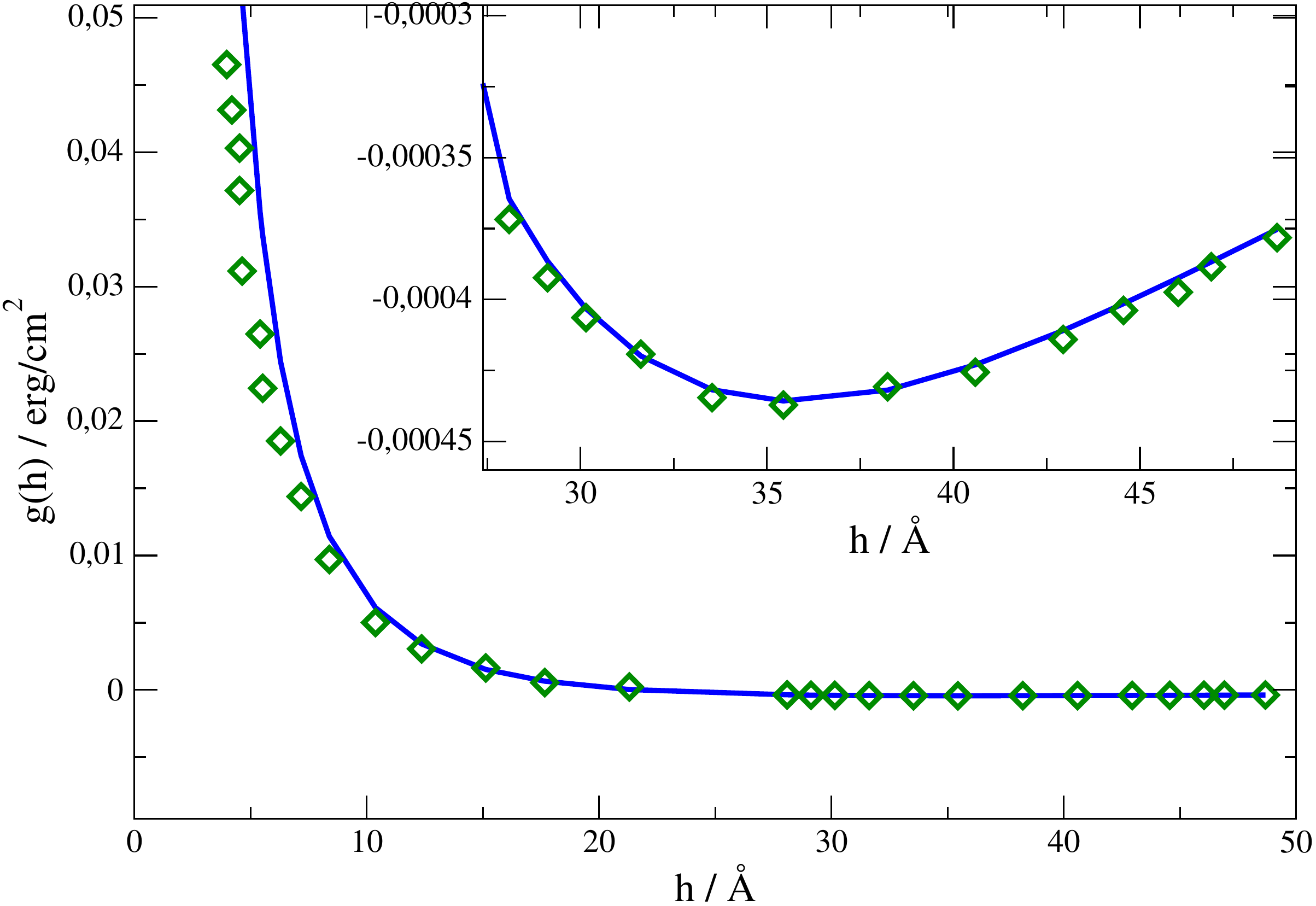}
        \caption{
                 \label{fig:ghvdw}
Van der Waals interface potential, as calculated numerically by Elbaum and
Schick (symbols), compared with the analytical approximation in {Supplementary} \Eq{crossover2}.
           }
\end{figure}

{Supplementary Fig.}~\ref{fig:ghvdw} shows a comparison of the exact results from Lifshitz
theory together with the fit to {Supplementary} \Eq{crossover2}, showing excellent agreement for
the set of parameters displayed in {Supplementary Table~1}. 
Since we find that {$g_\textrm{vdw}(\hh)$} is a factor of 1/100 smaller than {$g_\textrm{sr}(\hh)$} in the range $\hh < 10$~\AA, the van der Waals forces therefore only become relevant at large distances, where {$g_\textrm{sr}(\hh)$} becomes negligible due to the exponential decay form that it has.

\section*{Supplementary Note 3: Fit to the interface potential}

 The computer simulation results for the interface potential are fitted
 to the expression {$g(\hh) = g_{\textrm{sr}}(\hh) + g_\textrm{vdw}(\hh)$}, with
 {$g_\textrm{sr}(\hh)$}, the structural short range contribution:
 {\begin{equation}\label{eq:short-range2}
       g_\textrm{sr}(\hh) = C_2 \exp(-\kappa_2 \hh) - C_1 \exp(- \kappa_1 \hh)
	 \cos(q_0 \hh + \alpha)
  \end{equation}}
We use the coefficients $C_i$,  $\kappa_2$, $\kappa_1$, $q_0$ and $\alpha$ as fitting parameters, setting $\kappa_2=2\kappa_1$, for simplicity. Since the interface potential obtained from simulation is exact up to an additive constant, we seek parameters by minimizing the least square deviations from the corresponding disjoining pressure $\Pi(h)=-\partial_hg(h)$. We include also a constraint in the minimization to force the minimum of the interface potential to be at {$g_\textrm{min}=-5.9\times 10^{-5}$~J/m$^2$}, consistent with the observed contact angle of a droplet on an $\alpha$ film. The parameter values obtained from this fitting 
may be found in the {Supplementary Table~1}. 
The value found for $q_0$ is consistent with a strong renormalization away from the value one would expect from mean field theory \cite{chernov88, evans92, henderson94}.
{
In our fits we find that the target depth of the primary minimum of the
interface potential could not be reached. Since the separation between the 
kinetic phase lines is dictated by the depth and free energy difference
of the minima, the phase lines in {Fig.~4 (main text)} appear very close to each other.
To illustrate the role of the well depth in separating the phase lines,
{Supplementary Fig.~1} uses an interface potential blown up by a factor of 30.
The resulting phase lines very much resemble the kinetic phase diagram
observed experimentally \cite{asakawa16}. A recent study indicates that
the van der Waals forces estimated by Elbaum and Schick used here suffer from
insufficient optical data and predict interactions that are one order
of magnitude too weak  \cite{luengo20,fiedler20}. This explains why the  energy minima
of our model interface potential are so shallow and produce such a small
separation between the kinetic phase lines.
}

\section*{Supplementary Note 4: Stochastic dynamics of the Sine Gordon + Capillary Wave model}

{

Consider a  microscopic realization of the premelting film  at a given instant,
obtained with atomistic detail e.g.\ in a Molecular Dynamics simulation. The detailed
state of this system, as given by the atomic positions, may be conveniently
described by two interface profiles {$\mhsf(\xx,t)$} and
{$\mhfv(\xx,t)$} of the solid/liquid and liquid/vapor interfaces by using a
suitable coarse-graining on the scale of a few molecular
diameters \cite{benet16,benet19,llombart20,llombart20b}. Thus, these
profiles are still microscopic scale quantities.

In the spirit of the Langevin equation, we expect that the time evolution
of the interface profiles will be dictated by two different processes.
First, a deterministic evolution that is driven by a coarse-grained
Hamiltonian, {$\mathcal{H}[\mhsf,\mhfv]$}. 
Second, a Gaussian random evolution that
describes the thermal fluctuations of the coarse-grained degrees of freedom.
Such a dynamical equation for a thin liquid film on an inert substrate was derived
in \cite{grun06}. Here, using heuristic arguments we generalise this to the present
case of a water film on ice with evaporation/condensation and freezing/melting.

The time evolution at the solid/liquid surface is the result of freezing and
melting events, which can be described by a non-conserved dynamics of
{$\mhsf(\xx,t)$} as:
{\begin{equation}\label{eq:ev1}
    \frac{\partial \mhsf}{\partial t} = \displaystyle
    -k{\sf}\frac{\delta \mathcal{H}}{\delta \mhsf} + R_{\sf}(\xx,t)
\end{equation} }
where {$R_{\sf}(\xx,t)$} is a white noise field that accounts 
for microscopic detailed balance at equilibrium.

Similarly, the time evolution at the liquid/vapor surface is the result of
condensation and evaporation of the liquid film, which is conventionally
described in terms of a non-conserved dynamics. However, since
the premelting film is fluid, we must also account for the spreading
dynamics of the film, which we can describe using the thin film
approximation. Taking together the two processes, this leads to
a generalization of the stochastic thin film equation in the presence
of evaporation/condensation:
{\begin{equation}\label{eq:ev2}
    \frac{\partial \mhfv}{\partial t} = \displaystyle
          \left[\nabla\cdot\frac{\mhh^3}{3\eta}\nabla  -k{\fv}\right ]
	    \frac{\delta \mathcal{H}}{\delta \mhfv} + R_{\fv}(\xx,t)
\end{equation} }
where we have introduced {$\mhh=\mhfv-\mhsf$} for short, while
{$R_{\fv}(\xx,t)$}  is a noise field that accounts
for random stress fluctuations within the premelting film, together with
random evaporation and condensation events at the liquid/vapor surface \cite{grun06}.
Notice the model assumes the lubrication approximation for the advective 
dynamics of
the thin liquid film, which is accurate provided the characteristic wavelength
of the lateral height variations is larger than the thickness of the liquid
layer. This condition is obeyed for the very small contact angle droplets
which is certainly the case for the system of interest here.

Of course, the evolution of the two surfaces is not independent, and
leads to a pair of coupled stochastic differential equations for the
dynamics of the premelting film:
{\begin{equation}\label{eq:sdyn}
\frac{\partial \mhsf}{\partial t} = 
-k{\sf}\frac{\delta \mathcal{H}}{\delta \mhsf} +  \sqrt{2k_\textrm{B}T k\sf}\,\xi_{\sf}(\xx,t),
\end{equation}
\begin{equation}\label{eq:sdyn_2}
\frac{\partial \mhfv}{\partial t} =  
\nabla\cdot
\left[\frac{\mhh^3}{3\eta}\nabla\frac{\delta \mathcal{H}}{\delta \mhfv} 
 + \sqrt{\frac{2 k_\textrm{B}T \mhh^3}{\eta}} \, \xi_\textrm{tf}(\xx,t)
\right]
-k{\fv}\frac{\delta \mathcal{H}}{\delta \mhfv} +  \sqrt{2k_\textrm{B}T k\fv}\,\xi_{\fv}(\xx,t)
- \frac{\Delta\rho}{\rho\liq}    
\frac{\partial \mhsf}{\partial t}.
\end{equation}
}
Notice that, on account of the premelting film's
incompressibility, changes in {$\mhsf$} are conveyed into {$\mhfv$}, so
that the full dynamics of {$\mhfv$} is dictated both by
condensation/evaporation and freezing/melting rates.
The stochastic nature of the growth process is described by
spatially and temporal uncorrelated white noise 
fields, {$\xi_{\sf}$}, {$\xi_{\fv}$} and {$\xi_\textrm{tf}$}
that describe
coarse-grained thermal fluctuations at the solid/liquid
surface, the liquid/vapor surface and the premelting film, respectively. 
Finally, the amplitude of the random noise is chosen such that linearized forms of
 \Eq{ev1} and \Eq{ev2} at equilibrium satisfy the fluctuation-dissipation
 theorem exactly (i.e.\ obey detailed balance).

In order to be more specific, we now consider an explicit form for the
Hamiltonian, based on the sine Gordon model for the description of the
solid/liquid surface, and the capillary wave Hamiltonian for the description
of the liquid/vapor surface \cite{benet16,benet19,llombart20b}:
{\begin{align}\label{eq:hsgcw}
\mathcal{H}[\mhsf,\mhfv]  =  \int \left[\frac{\gamma{\sl}}{2}(\nabla \mhsf
)^2+\frac{\gamma{\lv}}{2}(\nabla \mhfv)^2-u\cos(q_z\mhsf) 
  + g(\mhfv-\mhsf) -\Delta p{\sl}\mhsf -\Delta p{\lv}\mhfv 
 \vphantom{\frac{\gamma{\sl}}{2}} \right]{\mathrm d}\mathbf{\xx},
\end{align}}
where {$\gamma{\sl}$}, the solid/liquid stiffness coefficient and {$\gamma{\lv}$}, 
the water/vapor surface tension  penalize the increase of surface area; the
cosine term favors solid/liquid film heights that are congruent with
the crystal lattice spacing as dictated by the wave-vector $q_z$; $u$
dictates the energy cost for excursions away from the preferred spacing;
{$g(\mhfv-\mhsf)$} is the
interface potential, which sets the equilibrium film height at coexistence;
and finally, {$\Delta p{\sl} = p_{\sol} - p_{\liq}$ and $\Delta p{\lv} = p_{\liq} -
p_{\vap}$, with $p_{\alpha}$} the bulk pressure of phase $\alpha$ are fields which account 
for the free energy cost of forming a liquid film at the expense of solid and vapor phases, respectively.
Notice however that the coefficients of this Hamiltonian are `bare' or mean field
parameters obtained from a microscopic theory averaged on the scale of the 
bulk correlation length.

Using the above Hamiltonian together with {Eqs.~\eqref{eq:sdyn} and \eqref{eq:sdyn_2}}, we obtain the following
explicit equation for 
the stochastic evolution of the coupled sine Gordon + Capillary Wave model:
{\begin{equation}\label{eq:SPDEs}
\frac{\partial \mhsf}{\partial t}  =  -k{\sf}[ \gamma\sf\nabla^2\mhsf 
 + w \sin(q_z \mhsf) - \phi\sl ] + \sqrt{2k_BT k\sf}\xi_{\sf}(\xx,t),
\end{equation}
\begin{equation}\label{eq:SPDEs_2} 
\frac{\partial \mhfv}{\partial t}  = 
 (\nabla \cdot \frac{\mhh^3}{3\eta}\nabla + k{\fv})[  \gamma\fv\nabla^2\mhfv + \phi\lv ] 
+ \nabla \cdot  \sqrt{\frac{2 k_BT \mhh^3}{3 \eta}} \xi_\textrm{tf}(\xx,t) +
   \sqrt{2k_BT k\fv}\xi_{\fv}(\xx,t)
      \displaystyle -\frac{\Delta\rho}{\rho\liq}\frac{\partial \mhsf}{\partial t},
\end{equation}
}
where $w=q_z u$, {$\phi\sl=\Delta p\sl -\Pi$} and {$\phi\lv=\Delta p\lv +\Pi$},
with $\Pi$ the disjoining pressure, defined as $\Pi(\hh) = -d g(\hh)/ d\hh$.

This result may be considered as a generalized stochastic thin film
equation \cite{davidovitch05,mecke05,grun06} that accounts also for variations of the underlying substrate by
means of the  sine Gordon model \cite{chui78,buttiker79,saito80}, and condensation/evaporation
by means of a  growth equation for rough surfaces \cite{karma93}. For
inert substrates and non volatile liquids ({$k\sf=k\fv=0$}), it reduces exactly to the 
stochastic thin film equation \cite{davidovitch05,mecke05,grun06}. For the buried
solid substrate below an infinitely thick liquid film in equilibrium ({$\mhfv=D$}, with
the constant $D\to\infty$), it reduces exactly to the stochastic sine Gordon equation
\cite{chui78,buttiker79,saito80,bennett81} and for an infinitely viscous  premelting film
($\eta=\infty$) under a flat inert substrate ({$k\sf=0$})
it recovers the stochastic growth equation of rough interfaces \cite{karma93}.
In each of these limiting cases, the amplitude of the noise is set such that
the fluctuation-dissipation theorem is obeyed exactly.

Obviously, the model does not incorporate any effects related to thermal
gradients. However, it is believed that for films less than approximately
$100$~nm, disjoining pressure effects largely dominate over thermo-capillary
forces \cite{pototsky05}. 
Furthermore, the experiments we describe appear to fulfil
local thermal equilibrium, since growth and evaporation events appear to 
be reversible and reproducible \cite{murata16}. 

Within this premise, the above result {incorporates many details of the
physics}. For equilibrium systems, with no forcing,
the stochastic dissipative equations serves as a starting point for
renormalization of the bare Hamiltonian \cite{saito80,nozieres87,moro01b}. 
For systems out
of equilibrium, it describes correctly purely dissipative
processes that occur deterministically when very large free energy gradients are
present.  Thanks to the stochastic term, it can also describe excursions away from the 
deterministic path when thermal fluctuations are comparable to the energy gradients,
and is also able to describe uphill activated processes against
the free energy gradients. In particular, the stochastic thin film equation
is able to predict the nucleation of metastable thin films \cite{grun06},
while the stochastic sine Gordon model can describe terrace nucleation
and activated crystal growth of faceted surfaces
\cite{buttiker79,bennett81}, as well as spiral growth \cite{karma98}.

Unfortunately, this detailed stochastic description can be in practice rather cumbersome,
as very lengthy simulations are required to observe activated processes,
while the stochastic nature of the dynamics implies the need to collect
averages over a large ensemble of trajectories. 

For this reason, it is convenient to perform an average of {Eqs.~\eqref{eq:SPDEs} and \eqref{eq:SPDEs_2}} over
the set of all random realizations of the noise subject to a given initial
condition, in a manner similar to that performed in Dynamical Density 
Functional Theory \cite{marconi99,archer04}. The reward for this additional
averaging is that the resulting evolution equation becomes deterministic,
and we can then avoid studying a large number of trajectories and
implementing the cumbersome details of stochastic differential equations.

To see this, assume
that {$\hsf(\xx,t)$} and {$\hfv(\xx,t)$} are the noise averaged film profiles.
Then, a given realization of the stochastic evolution as described
by {Eqs.~\eqref{eq:SPDEs} and \eqref{eq:SPDEs_2}} may be expressed in terms of deviations away from the
averaged profile as {$\mhsf(\xx,t) = \hsf(\xx,t) + \delta
\mhsf(\xx,t)$}, and likewise for {$\mhfv(\xx,t)$}.  Plugging these
into {Eqs.~\eqref{eq:SPDEs} and \eqref{eq:SPDEs_2}},
and performing an average over all realizations of the noise,
linear terms in {$\mhsf$} are immediately transformed into terms with
exactly the same form in {$\hsf$},
by definition, while the random noise terms vanish.
Therefore:
{\begin{equation}\label{eq:PDEs}  
\frac{\partial \hsf}{\partial t}  =  -k{\sf}[ \gamma\sf\nabla^2\hsf 
+ w \langle \sin(q_z \mhsf) \rangle_{\xi} -  \langle \phi\sl \rangle_{\xi} ],
\end{equation}
\begin{equation}\label{eq:PDEs_2} 
\frac{\partial \hfv}{\partial t}  = 
 \langle (\nabla \cdot \frac{\mhh^3}{3\eta}\nabla + k{\fv})[
    \gamma\fv\nabla^2\mhfv
 + \phi\lv ] \rangle_{\xi}
      \displaystyle -\frac{\Delta\rho}{\rho\liq}\frac{\partial \hsf}{\partial t},
\end{equation}}
where $\langle \rangle_{\xi}$ stands here for the average over the 
random trajectories. In the spirit of the Smoluchowski equation, the 
ensemble average over trajectories can be replaced by an ensemble
average over the film fluctuations, {$\delta \mhsf(\xx,t)$} and 
{$\delta \mhfv(\xx,t)$}. Then, assuming local thermal equilibrium
during the time evolution of the averaged profile,  the
random film height deviations {$\delta \mhsf(\xx,t)$} and
{$\delta \mhfv(\xx,t)$} are Gaussian distributed, with
a variance that is given by the equilibrium thermal fluctuations.
Accordingly, the ensemble over trajectories may be replaced by
a canonical ensemble over fluctuations  $\langle \rangle_{\xi}\to \langle
\rangle_T$.

In this approximation, we can readily interpret the first
of the coupled equations. The exact Gaussian average of
{$\sin(q_z \mhsf)$} readily yields:
{\begin{equation}
  \langle  \sin(q_z \mhsf) \rangle_T = e^{-\frac{1}{2}q_z^2 \langle \delta \mhfv^2 \rangle_T}
                     \sin(q_z \hsf).
\end{equation} }
Accordingly, the sinusoidal term {$w \sin q_z \mhsf$} of the original Hamiltonian, that
is a function of the microscopic film profile {$\mhsf$}, is transformed exactly
into a sinusoidal term that is a function of the averaged film profile {$\hsf$}, albeit with a
renormalized amplitude
{$w e^{-\frac{1}{2}q_z^2 \langle \delta \mhfv^2 \rangle}$}
\cite{saito80,prestipino99,moro01}.  At a higher level of approximation,
also the surface tension {$\gamma\sf$} is renormalized \cite{chui78,nozieres87},
but this difficulty need not concern us here since this level of
renormalization predicts the location of the roughening transition exactly.

On the other hand, the Gaussian average of {$\phi_{\sl}(\mhh)=\Delta p\sl -
\Pi(\mhh)$} yields:
{\begin{equation}
   \langle \phi_{\sl}(\mhh) \rangle_T = \Delta p\sl - \langle \Pi(\mhh)\rangle_T  
\end{equation} }
where $\langle \Pi(\mhh)\rangle$ is the Gaussian renormalized disjoining
pressure. In our model, the disjoining pressure consists of
short ranged (exponentially decaying) terms, and an algebraically decaying term.  The latter
does not renormalize, because of the long range nature, and need not concern
us any longer \cite{dietrich85} while the former  can be worked out
exactly under Gaussian renormalization \cite{chernov88,henderson94}. The result  
is again formally equal to the bare disjoining pressure of the Hamiltonian, 
albeit with renormalized coefficients, as discussed in the main text.

It follows that the dynamics of {$\hsf$} may be cast as:
{\begin{equation}\label{eq:evm1}
   \displaystyle   
\frac{\partial \hsf}{\partial t}  =  -k{\sf}[ \gamma\sf\nabla^2\hsf 
+ w_\textrm{R} \sin(q_z \hsf)  -  \Delta p\sl + \Pi_\textrm{R}(\hh) ] 
\end{equation} }
where the subscript {$\textrm{R}$} stands for Gaussian renormalized quantities.
This equation is formally {identical to \Eq{SPDEs}},
albeit with the microscopic film heights replaced by average film heights,
and the bare parameters of the Hamiltonian replaced by
renormalized coefficients.

A similar result for the time evolution of {$\hfv$} cannot be readily
obtained, because the mobility coefficient in the lubrication approximation
depends on the film thickness.
Hence, the Gaussian average of the Hamiltonian couples with  the $\hat\hh^3$
term in the mobility coefficient.
However,  in our system
the fluctuations of {$\mhsf$} are smaller than one lattice
spacing because the surface
is smooth, while the fluctuations of {$\mhfv$} are limited by the
long range van der Waals tail and increase logarithmically with the film height.
Accordingly, we expect that the fluctuations of $\mhh$ away from the mean
value $\hh$ will be small.
This allows us to expand the mobility in powers of $\delta \mhh$ and
retain only the leading order term. The time evolution of {$\hfv$}
is then given as:
{\begin{equation}\label{eq:evm2}
   \displaystyle   
\frac{\partial \hfv}{\partial t}  = 
 (\nabla \cdot \frac{\hh^3}{3\eta}\nabla + k{\fv})[
    \gamma\fv\nabla^2\hfv
 +  \Delta p\lv + \Pi_R(\hh) ]
\end{equation} }
where again the subscript {$\textrm{R}$} stands for Gaussian renormalized quantities.

\Eq{evm1} and \Eq{evm2} provide a system of deterministic coupled differential
equations for the average dynamics of the coupled stochastic sine Gordon and
stochastic thin film equations:
{\begin{equation}\label{eq:DPDEs}  
\frac{\partial \hsf}{\partial t}  =  -k{\sf}[ \gamma\sf\nabla^2\hsf 
 + w \sin(q_z \hsf) - \phi\sl ],
\end{equation}
\begin{equation}\label{eq:DPDEs_2}  
\frac{\partial \hfv}{\partial t}  = 
 (\nabla \cdot \frac{h^3}{3\eta}\nabla + k{\fv})[  \gamma\fv\nabla^2\hfv + \phi\lv ] 
      \displaystyle -\frac{\Delta\rho}{\rho\liq}\frac{\partial \hsf}{\partial t}.
\end{equation}}
From {Eqs.~\eqref{eq:DPDEs} and \eqref{eq:DPDEs_2}} we can readily see that the trajectory
averaged film profiles {$\hsf$} and {$\hfv$} follow a time evolution
that is exactly the same as that for the microscopic profiles {$\mhsf$} and {$\mhfv$}
given in {Eqs.~\eqref{eq:SPDEs} and \eqref{eq:SPDEs_2}},
albeit with renormalized parameters of the Hamiltonian.

In view of this, we can write the coupled time evolution of {$\hsf$} and {$\hfv$}
as in Eq.~(3) of the main manuscript, with the understanding that 
the free energy, $\Omega$ is a Gaussian renormalized Hamiltonian
$\Omega=\langle \mathcal{H}\rangle_{\xi}\approx \langle \mathcal{H}\rangle_{T}$, which adopts the same form
as the mean field Hamiltonian, \Eq{hsgcw}, albeit with renormalized coefficients.

This interpretation is very much analogous to the similar results found in
Dynamic Density Functional Theory \cite{marconi99,archer04,lutsko12},
which is a theory for the dynamics of the density distribution of Brownian
stochastic particles (i.e.\ colloids). There, the resulting deterministic equation is
an evolution equation for the average density profile, obtained by averaging over
all realizations of the noise, while the stochastic equation describes the evolution of the
microscopic density for
one particular realization of the noise. The input to the deterministic equation is the free
energy functional known from equilibrium Density Functional Theory \cite{evans92}.
In this picture, the
deterministic evolution of the microscopic density can also be interpreted
to describe the most likely path of the stochastic process when the
fluctuations are small \cite{lutsko12,lutsko19}. In this case, the microscopic Hamiltonian
does not deviate significantly from the renormalized free energy, and then
the average evolution is essentially the same as that of the most likely path,
as expected in mean field.

}

\section*{Supplementary Note 5: Mean field dynamics and kinetic phase diagram}

The dynamics of the premelting film, i.e.\ of the solid/liquid and liquid/vapor
interfaces {$\hsf$} and {$\hfv$} respectively, is governed by the free energy in Eq.~(3) of the main text, together with the gradient dynamics equations in Eq.~(4) of the main text. 

The dynamics exhibited by this pair of coupled partial differential equations is
very rich, and the full gamut can only be found by solving numerically. However,
analytic results can be obtained for the long-time average behavior, i.e.\ for
the growth speeds. For {$\phi\sl^2<w^2$}, ice growth (corresponding to {$L\sl$}
increasing) cannot occur, because the thermodynamic force {$\phi\sl$} is not
sufficient to overcome the sinusoidal pinning potential. Therefore, growth
proceeds by the horizontal spread  of terraces with velocity {$\frac{\pi}{4} k\sf
(\frac{\gamma\sl}{u})^{1/2} \phi\sl$} \cite{saito80,nozieres87}. For
{$\phi\sl^2>w^2$}, the driving potential {$\phi\sl$} overcomes the sinusoidal
potential, and uniform growth can occur. However, if {$\phi\sl$} is only
marginally larger than $w$, the process occurs in a stepwise fashion, with a
long interval in which there is almost no growth, followed by fast growth over a
short time period, as observed in computer simulations \cite{neshyba16}. This
leads to a height increment of $\approx 2\pi/q_z$, i.e.\ of one ice lattice
spacing. The process repeats recursively with a period {$\tau=2\pi/q_z\sqrt{\phi\sl^2-w^2}$}, so that the average growth rate is {$k\sf\sqrt{\phi\sl^2-w^2}$}. For large {$\phi\sl$}, this provides the usual `linear growth' mode of rough interfaces, but in the limit {$\phi\sl\approx w$}, the linear growth mode can be much slower than the horizontal translation of terraces.

For flat films, the average growth rate over time scales much larger than $\tau$ is then given by
{\begin{equation}\label{eq:mfd}
\langle\frac{\partial \hsf}{\partial t}\rangle  =  \pm k{\sf} \sqrt{\phi\sl^2-w^2}
\end{equation}
\begin{equation}\label{eq:mfd_2}
   \displaystyle   
\langle\frac{\partial \hfv}{\partial t}\rangle  = 
 k{\fv} \phi\lv -\frac{\Delta\rho}{\rho\liq}\langle\frac{\partial \hsf}{\partial t}\rangle
\end{equation}}
where the plus sign stands for freezing, and the minus sign for sublimation. Subtracting one from the other, we obtain the average speed of the liquid film thickness growth
{\begin{equation}
   \displaystyle \langle\frac{\partial \hh}{\partial t}\rangle  =  
  \displaystyle  k\fv \phi\lv \mp \frac{\rho\sol}{\rho\liq} k\sf
\sqrt{\phi{\sl}^2-w^2}.
\end{equation}  }
This result becomes particularly simple for the case when $w=0$, as discussed in
the text. In the general case where $w\neq0$ and {$p>p\sv$} so that the height of the ice grows, the condition that the liquid thickness is stationary {$\langle{\partial_t \hh}\rangle=0$} is achieved for {$\phi\lv\ge 0$}, {$\phi\sl\ge 0$}, and  {$\phi\sl^2-w^2\ge 0$}. In the marginal case where {$\phi\sl=w$}, then we need {$\phi\lv=0$}. Solving these two conditions simultaneously corresponds to {$\Delta p\sl + \Delta p\lv = \pm w$}. Using the approximate but nonetheless accurate thermodynamic relations for the pressure differences given below in Eqs.~(\ref{eq:laplacelv}) and (\ref{eq:laplaceiv}), these condition may be solved as a function of $T$, yielding the following equation for the boundary 
{{\begin{equation}
  p_\textrm{ns}(T) = p\sv(T) e^{\pm \frac{w}{\rho\sol k_\textrm{B}T}}.
\end{equation} }}
States between the sublimation line {$p\sv(T)$} and the boundary line {$p_\textrm{ns}(T)$} neither grow nor sublimate because the surface {$\hsf$} can not grow in the absence of thermal activation.

For the more general case when {$\phi\sl^2-w^2\ge 0$}, the stationarity condition is achieved as a solution of the equation
{\begin{equation}\label{eq:stationary_condition}
  \displaystyle  k\fv \phi\lv \mp \frac{\rho\sol}{\rho\liq} k\sf
  \sqrt{\phi\sl^2 - w^2} = 0.
\end{equation} }
It corresponds to the condition that the liquid/vapor and solid/liquid surfaces grow at the same rate. Only one solution exists, given that the surface growth rates are monotonic. However, in order to solve explicitly we need to square each term. The resulting equation then has two solutions, each of the same magnitude but with opposite sign. Of course, one is unphysical. Therefore, squaring in \Eq{stationary_condition} we obtain
{\begin{equation}\label{eq:stationarity2}
  \rho\sol^2 k\sf^2\phi\sf^2 - \rho\liq^2 k\fv^2 \phi\lv^2 =
 \rho\sol^2 k\sf^2 w^2 ,
\end{equation} }
under the condition that {$p\vap>p\sl(T)$}. This provides a quadratic equation for $\Pi$ as a function of {$p\vap$} and $T$, so one obtains
{\begin{equation}
\Pi = -\Delta p_\textrm{k},
\end{equation} }
with
{\begin{equation}\label{eq:solve_stat2_1}
\Delta p_\textrm{k} = - \frac{f\sol^2\Delta p\sl+f\liq^2\Delta p\lv\pm
  f\sol \left[ f\liq^2(\Delta p\sl + \Delta p\lv)^2 + (f\sol^2-f\liq^2) w^2    \right]^{1/2}
}{(f\sol^2-f\liq^2)},
\end{equation} }
where {$f\sol=\rho\sol k\sf$} and {$f\liq =\rho\liq k\fv$}. Thus, the solution may formally be written in exactly the same form as the equilibrium condition for the adsorption on an inert substrate, with the Laplace pressure difference {$\Delta p = p_\textrm{l} - p_\textrm{v}$} replaced by a kinetic pressure difference $\Delta p_{\rm k}$ which depends on the growth mechanism and rate constants. Likewise, an effective potential exists whose extrema are stationary states of the underlying dynamics.

Alternatively, \Eq{stationarity2} may be solved for {$p\vap$} as a function of $\Pi$ and $T$, with the result:
{\begin{equation}\label{eq:solve_stat2_2}
\rho\liq k_\textrm{B}T\ln\frac{p}{p\lv} + \Pi = -
\frac{\Delta\rho C \pm \left [ \kappa^2\rho\liq^2 C^2 + \rho\liq^2 w^2
(\Delta\rho^2 - \kappa^2\rho\liq^2) \right ]^{1/2} }{\Delta\rho^2-\kappa^2\rho\liq^2},
\end{equation} }
where {$\kappa=\rho\liq k\fv/\rho\sol k\sf$} and {$C=\rho\sol\rho\liq
k_\textrm{B}T\ln\frac{p\fv}{p\sv} - \rho\sol\Pi$}. In this case, the result corresponding
to $w=0$ and $\Pi=0$ (for a rough ice surface) is obtained for the `+' root.
Three kinetic transition lines in the phase diagram can be identified from the
numerical solution of these equations as discussed in the main text.
Particularly, the kinetic coexistence line between $\alpha$ and $\beta$ states
obeys a double tangent construction of wetting phase diagrams, albeit with the
kinetic overpressure replacing the equilibrium value:
{\begin{eqnarray}
   \displaystyle \omega_\textrm{k}(h_1)  & = &  \omega_\textrm{k}(h_2) \\
   \displaystyle \Pi(h_1) & = & -\Delta p_\textrm{k} \\
   \displaystyle \Pi(h_2) & = & -\Delta p_\textrm{k}
\end{eqnarray}}
The first condition imposes equal effective free energy for both films, and the
other two impose that both states obey the quasi-stationary condition at equal
kinetic overpressure {$-\Delta p_\textrm{k}$}. Alternatively, these equations may be written more concisely as:
{\begin{eqnarray}
   \displaystyle g(h_1) + \Pi(h_1) h_1  & = &  g(h_2) + \Pi(h_2) h_2 \\
   \displaystyle \Pi(h_1) & = & \Pi(h_2)
\end{eqnarray}}
Once the value of $\Pi$ that satisfies the condition is known, the pressure {$p\vap$} at which the condition is met can be obtained by solving \Eq{stationarity2} for {$p\vap(T)$} using the appropriate value of $\Pi$ in \Eq{solve_stat2_2}.

From these observations we are able to construct the highly detailed kinetic phase-diagram shown in Fig.~4 of the main text, which is an essential tool for understanding at different state points the numerical results obtained from the coupled gradient dynamics partial differential equations in Eq.~(4) of the main text.

\section*{Supplementary Note 6: Thermodynamic functions and the equilibrium phase diagram}

The pressure differences between solid/liquid and liquid/vapor phases are the thermodynamic driving forces that lead to the growth of the ice and the liquid from the vapor. To determine these differences requires knowledge of the equilibrium phase diagram, i.e.\ to know the pressure as a function of temperature along the condensation and sublimation lines, {$p\sl(T)$} and {$p\sv(T)$}, respectively. We obtain these by assuming they follow from the Clausius-Clapeyron equation. This approximation is excellent for the sublimation line \cite{feistel07}, and remains good for the vaporization line down to 260~K \cite{murphy05}. They are given by
{ \begin{equation}\label{eq:ccice}
     \ln \frac{p\sv}{p_\textrm{t}} = \frac{\Delta H{\sv}}{R} \left( \frac{1}{T_\textrm{t}} - \frac{1}{T} \right),
  \end{equation}
  \begin{equation}\label{eq:ccwater}
       \ln \frac{p\lv}{p_\textrm{t}} = \frac{\Delta H{\lv}}{R} \left( \frac{1}{T_\textrm{t}} - \frac{1}{T} \right),
   \end{equation}}
where {$T_\textrm{t}$} and {$p_\textrm{t}$} are the temperature and pressure at the triple point, $R$ is the gas constant, {$\Delta H\sv$} is the molar enthalpy change for sublimation and {$\Delta H\lv$} is the molar enthalpy change for condensation.

Since ice and water can both be treated as effectively being incompressible, the pressure changes which are relevant to this study are very small. Therefore, the pressure differences can accurately be approximated by
  { \begin{equation}\label{eq:laplacelv}
	  p\liq - p\vap = \rho\liq RT \ln \frac{p\vap}{p{\lv}(T)},
   \end{equation}
   \begin{equation}\label{eq:laplaceiv}
	    p\sol - p\vap = \rho\sol RT \ln \frac{p\vap}{p{\sv}(T)}.
   \end{equation}}
Using Eqs.~(\ref{eq:ccice})--(\ref{eq:laplaceiv}), we obtain explicit expressions for the liquid-vapor and ice-liquid overpressures as
{\begin{equation}\label{eq:plv}
		    p\liq - p\vap = \rho\liq RT \ln \frac{p\vap}{p_\textrm{t}} -
		    \frac{\rho\liq\Delta H{\lv}(T-T_\textrm{t})}{T_\textrm{t}},
 \end{equation}
 \begin{equation}\label{eq:psl}
		       p\sol - p\liq = (\rho\sol-\rho\liq) RT \ln \frac{p\vap}{p_\textrm{t}}
			 + \frac{(\rho\liq\Delta
			 H{\lv}-\rho\sol\Delta H{\sv})(T-T_\textrm{t})}{T_\textrm{t}}.
 \end{equation}}
Notice that the pressure difference between the solid and liquid phases
decreases as the ambient vapor pressure increases. The triple point data
required for the implementation of Eqs.~(\ref{eq:ccice})--(\ref{eq:psl}) may be
found in {Supplementary Table~2}.

\section*{Supplementary Note 7: Kinetic coefficients for the growth rate laws}

\subsubsection*{Growth of the liquid/vapor surface}

For an infinitely thick premelting film with a flat liquid-vapor surface, \Eq{DPDEs} for the growth rate of the surface becomes
{\begin{equation}
\frac{\partial \hfv}{\partial t}  = k{\fv} \Delta p\lv .
\end{equation} }
Replacing {$p{\liq}-p\vap\approx\rho\liq k_\textrm{B}T (p-p{\lv})/p{\lv}$} in the term for condensation/evaporation rate, we find
{\begin{equation}
   \frac{\partial \hfv}{\partial t} \approx k\fv \rho\liq k_\textrm{B}T (p-p{\lv})/p{\lv}.
\end{equation} }
This result can be compared to the Knudsen-Hertz law, which reads
{\begin{equation}
   \frac{\partial \hfv}{\partial t} = k_\textrm{KH} (p-p{\lv}),
\end{equation} }
where {$k_\textrm{KH}=\alpha\fv/\rho\liq (2\pi m_\textrm{w} k_BT)^{1/2}$}, and where {$\alpha\fv$} is the sticking coefficient, or fraction of vapor molecules that stick to the interface upon collision and {$m_\textrm{w}$} is the mass of a water molecule. Therefore, we find
{\begin{equation}
  k\fv = \frac{p{\lv}}{\rho\liq k_\textrm{B}T} k_\textrm{KH}.
\end{equation} }
We calculate {$k\fv$} using the thermodynamic data reported in 
{Supplementary Table~2.} 
We also assume {$\alpha\fv=1$} for the attachment of pure water vapor onto the ice surface, consistent with all current molecular simulation studies \cite{batista05, neshyba09, pfalzgraf11, llombart19}.

\subsubsection*{Growth of the solid/liquid surface}

For an infinitely thick premelting film with flat solid-liquid interface, \Eq{DPDEs} for the growth rate of the surface becomes
{\begin{equation}
\frac{\partial \hsf}{\partial t}  = k{\sf} \Delta p\sl .
\end{equation} }
Replacing {$p\sol-p\vap \approx \rho\sol \Delta H\sl \frac{T-T_\textrm{t}}{T_\textrm{t}}$} in the term for the freezing/melting rate we find
{\begin{equation}
   \frac{d \hsf}{dt} \approx k\sf \rho\sol \Delta H\sl \frac{T_\textrm{t}-T}{T_\textrm{t}}.
\end{equation} }
This result can be compared to the law of linear growth for a crystal from the
melt which holds at large undercooling \cite{pruppacher10},
{\begin{equation}
   \frac{d \hsf}{dt} = k_\textrm{LG} (T_\textrm{t}-T).
\end{equation} }
The result for the rate constant suggested by Librecht \cite{libbrecht14},
{$k_\textrm{LG}=0.07$}~cm/s~K, leads to {$k\sf=6\times 10^{-10}$}~m/s Pa. However, the
slope of the kinetic coexistence line is determined by the ratio {$k\sf/k\fv$},
and we find that the slopes observed in experiments can only be reproduced for {$k\sf/k\fv \approx 6.4$}, which is about a factor of 10 smaller. It seems likely that the kinetic coefficient for growth from the premelting film could be significantly smaller than that from the melt, since the interface is considerably smoother \cite{benet16}. Therefore, in our calculations we set {$k\sf=6.4 k\fv$}.

\subsection*{Size of the region where nucleated dynamics occurs}

For $\hh\to\infty$, our model gives an equation for the dynamics of {$\hsf$} that corresponds to the growth of ice within supercooled water. This is
{\begin{equation}
 \frac{\partial \hsf}{\partial t} = k\sf ( \gamma{\sl} \nabla^2 \hsf - u q_z \sin(q_z \hsf)  +   \Delta p\sl),
\end{equation} }
which is a forced overdamped sine-Gordon equation. The growth is nucleated for {$u q_z > \Delta p\sl$}, and otherwise linear in time \cite{buttiker81, bennett81}. Therefore, we can obtain an order of magnitude estimate for the parameter $u$ from the value of the temperature where there is a crossover from nucleated to linear growth of ice in supercooled water. According to Pruppacher \cite{pruppacher10}, this occurs at about {$T-T_\textrm{t}\approx2$~K}. Using {$p\sol-p\liq=\rho_{\sol}\Delta H_{\sl} \Delta T/T_\textrm{t}$}, we find
{\begin{equation}
      u = \frac{d_\textrm{B}}{2\pi} \rho{\sol} \Delta H{\sl} \frac{\Delta T^*}{T_\textrm{t}}.
\end{equation} }
Using $\Delta T^*=2$~K as suggested from results in Ref.~\cite{pruppacher67},
and {$d_\textrm{B}=0.37$}~nm, we find $u=1.3\times 10^{-4}$~J/m$^2$. This is about five
times larger than the results obtained from computer simulations, which yield
$u=2.8\times 10^{-5}$~J/m$^2$ \cite{benet16, benet19, llombart20b}. The value we
use is given in {Supplementary Table~3}. 

\subsection*{Viscosity}

In principle, the lubrication approximation on which our thin film dynamics model is based on uses as input the bulk liquid viscosity. Some studies suggest there is a large enhancement of the viscosity of premelting films (c.f.\ \cite{pramanik17}) over the bulk value. However, this appears to remain as an unsolved issue, with very recent high-profile studies being published \cite{PRX19gliding}. Thus, here we use the viscosity of supercooled bulk water as reported in 
{Supplementary Table~3}. 
Changing the value of the viscosity in our model will not qualitatively 
change our results.

Data for all the parameters used in the Singe Gordon + Capillary Wave dynamical
model may be found in {Supplementary Table~3}.

\section*{Supplementary Note 8: Numerical solution of the gradient dynamics} 

Numerical computations of the dynamics of the interfaces predicted by our
coupled partial differential equation model in Eqs.~(\ref{eq:DPDEs}) {and \eqref{eq:DPDEs_2}} (i.e.\ Eqs.~(3) and (4) of the main text) are performed using a method of lines technique similar to that used in Ref.~\cite{yin17}. The method is extended to evolve the two interfaces (solid-liquid, and liquid-vapor), with coupling terms involving mass transfer and the two interface potentials naturally included. However, we evaluate the spatial derivatives in a different manner, which significantly increases the rate of numerical convergence. This was done because for the evolution of the solid-liquid interface, a pinning effect in the horizontal direction can occur if too few mesh points are used. Consequently, rather than using an extremely large number of points in the finite difference scheme used in \cite{yin17}, here we implement a periodic pseudospectral method.

The numerical method uses results from Ref.~\cite{trefethen00}, discretising on a regular (periodic) grid and uses a band-limited interpolant derived using the discrete Fourier transform and its inverse to form the differentiation matrices which act in real space (see chapter 3 of \cite{trefethen00} for details). The periodicity enabled by the premelting film avoids the need to evolve actual contact lines, in comparison to some of our previous work using pseudospectral discretisation \cite{DNS_Shikh,DNS_Crack}. For the time stepping, the ode15s Matlab variable-step, variable-order solver is used \cite{shampine97}. Our numerical calculations are performed on the nondimensionalised version of the model equations. We find that choosing {$\kappa_1^{-1}\approx 0.49$~nm and $3\eta/(\kappa_1 \gamma_\textrm{lv})\approx 0.11$}~ns as our units of length and time in the nondimensionalisation works well.

To explore the effectiveness of our model to at least qualitatively reproduce the phenomena observed in the experiments and to confirm the validity of the analytical predictions for the different behaviors in the different {$(p_\textrm{v},T)$} regions of the phase diagram, we perform an extensive set of full numerical simulations, for a range of state points covering all the different growth regimes. Of course, the observed behavior also depends on the effective surface free energy {$\omega_\textrm{k}(h)$}, which includes ice surface effects on the evolution of the interfaces, and on the initial conditions. A comprehensively large variety of initial conditions (i.e.\ the $t=0$ profiles of the two interfaces) have also been trialled, especially for planar interfaces (at different separations, usually based on the heights corresponding to the $\alpha$ or $\beta$ minima) with either small imperfections in the solid, or an initial perturbation of the liquid surface, or both. The results presented in the paper are drawn from the following three different initial condition types: Firstly, a planar solid-liquid surface with a Gaussian droplet shaped perturbation in the liquid-vapor interface, given by
{\begin{align}
L_\textrm{lv} &= d_\textrm{B} + h_0 + A_\textrm{f}\exp[-((x-x_\textrm{L}/2)/x_\textrm{wf})^2],\\
L_\textrm{sl} &= d_\textrm{B}, 
\end{align}}
where $h_0$ is an initial separation (such as the height of the $\alpha$ minimum), {$x_\textrm{L}$} is the size of the periodic domain (taken as {$x_\textrm{L} =2500\kappa_1^{-1}$}) in all simulations presented here, {$A_\textrm{f}=17\kappa_1^{-1}$} is the height of the Gaussian perturbation and {$x_\textrm{wf}$} is a measure of its width. We typically set {$x_\textrm{wf}=450\kappa_1^{-1}$} for the results presented here.

The other two forms for the initial conditions are
{\begin{align}
L_\textrm{sl} &= d_\textrm{B}
\pm \frac{A_\textrm{i}}{2}d_\textrm{B} \left[\tanh\left(\frac{x-(x_\textrm{L}-x_\textrm{wi})/2}{10\kappa_1^{-1}}\right)-\tanh\left(\frac{x-(x_\textrm{L}+x_\textrm{wi})/2}{10\kappa_1^{-1}}\right)\right]
,\\
L_\textrm{lv} &= d_\textrm{B} + h_0, 
\end{align}}
which corresponds to a planar liquid-vapor surface, together with an ice-liquid interface that has on it a small imperfection of hight {$A_\textrm{i}$} that is an integer multiple of the height of a single ice terrace, that protrudes either into or away from the liquid, and has width {$x_\textrm{wi}$}. Values used in the work presented here are {$\{A_\textrm{i},x_\textrm{wi}\}=\{1,x_\textrm{wi}=x_\textrm{L}/16\}$} and {$\{A_\textrm{i},x_\textrm{wi}\}=\{10,x_\textrm{wi}=9x_\textrm{L}/16\}$}.

Fig.~5 of the main text displays snapshots from four typical simulations, and here we show snapshots from two additional simulations in Figs.~\ref{fig:s2}--\ref{fig:s4}. The full time evolutions of all six simulations can be seen in the movies included as supplementary material, named Movies S1--S6.

\section*{Supplementary References}

\end{document}